\documentclass[reprint,amsmath,amssymb,aps,pra,showpacs,floatfix,superscriptaddress,longbibliography]{revtex4-1}

\usepackage{graphicx}
\usepackage{bm}
\usepackage{hyperref}
\usepackage{verbatim}
\usepackage{color}
\usepackage[outdir=./]{epstopdf}

\renewcommand{\d}{{\rm d}}
\newcommand{\e}{{\rm e}}

\usepackage[normalem]{ulem}

\begin{document}

\title{Ignition of quantum cascade lasers in a state of oscillating electric field domains}

\author{David O. Winge}
\affiliation{Mathematical Physics and NanoLund,
  Lund University, Box 118, 22100 Lund, Sweden}
\affiliation{Department of Physics, University of California San Diego, La Jolla,  California 92093-0319, USA}
\author{Emmanuel Dupont}
\affiliation{Advanced Electronics and Photonics Research Centre, National Research Council, Ottawa, Ontario K1A 0R6, Canada}
\author{Andreas Wacker}
\email[corresponding author: ]{Andreas.Wacker@fysik.lu.se}
\affiliation{Mathematical Physics and NanoLund,
  Lund University, Box 118, 22100 Lund, Sweden}
\date{22 July 2018: Accepted manuscript, published in Phys.~Rev.~A {\bf 98}, 023834 (2018)}

\begin{abstract}
Quantum Cascade Lasers (QCLs) are generally designed to avoid negative
differential conductivity (NDC) in the vicinity of the operation point
in order to prevent instabilities.
We demonstrate, that the
threshold condition is possible under an inhomogeneous distribution
of the electric field (domains) and leads to lasing at an operation
point with a voltage bias normally attributed to the NDC region.
For our example, a
Terahertz QCL operating up to the current maximum temperature of 199
K, the theoretical findings agree well with the experimental
observations. In particular, we  experimentally observe self-sustained
oscillations with GHz frequency before and after threshold. These are
attributed to traveling domains by our simulations. Overcoming the
design paradigm to avoid NDC may allow for the further optimization of
QCLs with less dissipation due to stabilizing background current.
\end{abstract}

\maketitle

%%%%%%%%%%%%%%%%%%%%%%%%%%%%%%%%%%%%%%%%%%%%%%%%%%%%%%%%%%%%%%%%%%%%
\section{Introduction}
%{Introduction}

The core of any laser is a gain medium, which is pumped  sufficiently
strong, so that the gain overcomes the losses at threshold and the
strong lasing field ignites.  Typically, the medium is in a stationary
state before threshold (while more complicated behavior such as
relaxation oscillations may occur afterwards).  For electrically
pumped lasers, this implies electric stability, which includes the
avoidance of negative differential conductivity (NDC). This is, e.g.,
a longstanding issue for the application of dispersive gain (also
called Bloch gain) in semiconductor superlattice (SLs), where the gain
mechanism is intrinsically related to NDC
\cite{EsakiIBM1970,KtitorovSovPhysSolState1972}. In this context, it
was pointed out by Kroemer,
that once lasing is established, a stable operation
point is possible in SLs, as the lasing field changes the electric
behavior substantially \cite{KroemerArxiv2000}. In a region of NDC, a
homogeneous electric field distribution in the transport direction is
unstable and decays into regions with different electric fields,
called field domains, as originally studied for Gunn diodes
\cite{ShawBook1992}. For SLs, domain formation
\cite{EsakiPRL1974,WackerPhysRep2002,BonillaRepProgPhys2005} hindered the observation of
dispersive gain \cite{ShimadaPRL2003,SavvidisPRL2004} for a long time
and  no laser action has been realized  so far. Thus, avoiding NDC is
a common design paradigm of the related, technologically extremely
successful Quantum Cascade Laser (QCL)
\cite{FaistScience1994,FaistBook2013}, as already pointed out in the
precursory work~\cite{KazarinovSPS1971}. While NDC and the formation of stationary domains
have been observed and analyzed in some QCL structures
\cite{LuPRB2006,WienoldJAP2011,YasudaProc2013,DharSciRep2014}, they are commonly
considered to impede the desired lasing action.

In this work, we demonstrate, that the operation of a QCL is also
possible in the range of NDC, where the lasing field stabilizes the
operation point similar to the never realized idea for SLs
\cite{KroemerArxiv2000}. The key issue is the ignition of the lasing
field, which occurs in a state of oscillating field domains, as shown
both experimentally and by simulations. Our findings reveal an
alternative type of laser ignition with interesting nonlinear
behavior. Furthermore, lifting the paradigm of NDC in QCLs, may also
allow for new designs with higher performances.

\begin{figure*}[bt]
\includegraphics[width=\textwidth]{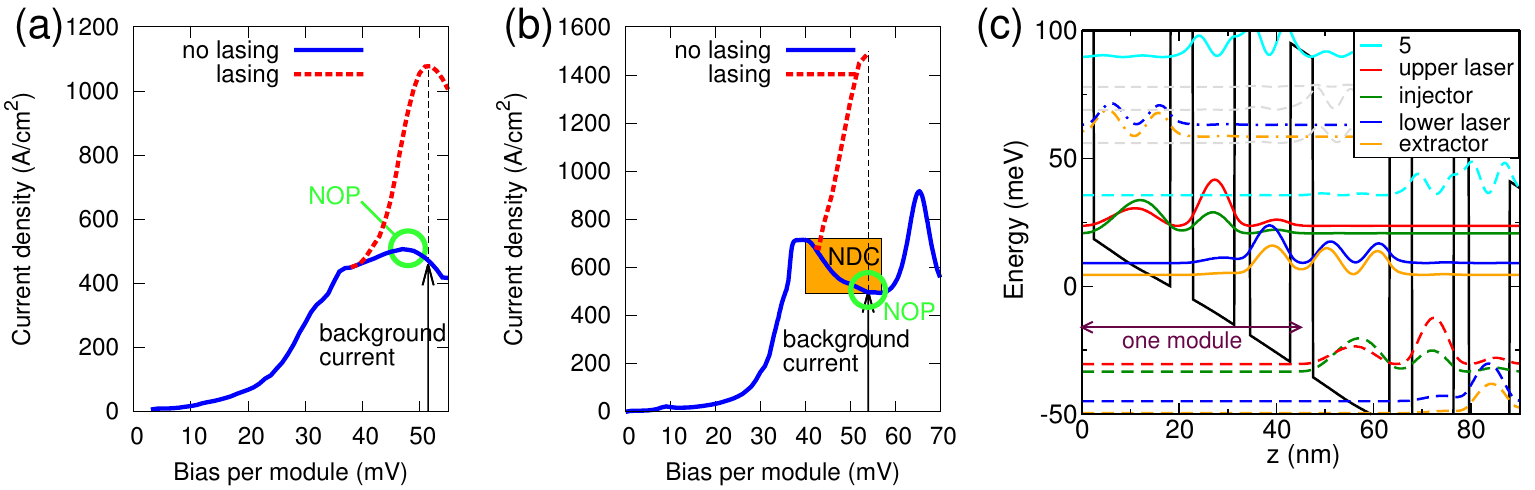}
\caption{(Color online) Operating in the NDC region enhances the ratio between
    the lasing driven current and the background current:  {\bf (a)}
  Calculated current versus bias assuming a homogeneous field
  distribution  for the sample \cite{LiElectronLett2014}.
  The green circle depicts the
  nominal operation point (NOP).  {\bf (b)} Corresponding result for
  the sample V812 \cite{FathololoumiJAP2013}, which is considered
  here. {\bf (c)} Conduction band profile and relevant
  electronic states (sequence between 0 and 50 meV as in legend  and periodically 
  shifted with different linestyles)  at the NOP of 54 mV/module for the sample V812
  where two modules are displayed. Starting from the extraction
  barrier, the layer sequence of a module is:
  \textbf{46}/158/\textbf{46}/86/\textbf{31.7}/83 \AA~where the
  boldface font stands for
  $\textrm{Al}_{0.15}\textrm{Ga}_{0.85}\textrm{As}$ barriers and roman
  font the GaAs wells. The center 5~nm of the 158~\AA~ injector well is
  Si-doped at $6\times 10^{16} /\textrm{cm}^3$. The entire device contains 222 modules.}
\label{FigHomogeneous}
\end{figure*}

Specifically, we consider a Terahertz QCL structure, labeled V812,
which was already studied in \cite{FathololoumiJAP2013}. It shows
lasing up to 199.3 K  (with Cu waveguides), comparable to the current
record of 199.5 K~\cite{FathololoumiOE2012} for a similar device with
higher current density. V812 belongs to the family of three-well
designs with resonant tunnel injection and resonant phonon
extraction~\cite{WilliamsAPL2003}, see the band-diagram at the nominal
operation point (NOP) in Fig.~\ref{FigHomogeneous}(c). This class of
designs has been previously shown to exhibit bias instabilities around
threshold~\cite{FathololoumiJAP2013}, when operated via a serial
resistance.  This is due to NDC occurring for biases after alignment
of the injector level with the lower laser and extractor level, as
common for tunneling in semiconductor heterostructures
\cite{KazarinovSPS1971,EsakiPRL1974,CapassoAPL1986}.

Fig.~\ref{FigHomogeneous}(a) shows the generally sought scenario where
the QCL ignites in a region without NDC; this example is a highly
performant bound-to-continuum four-well structure
\cite{LiElectronLett2014}. After threshold, the lasing field causes
stimulated emission, which enhances the current (dashed line) compared
to the current without lasing (solid line).  These data are obtained
from our non-equilibrium Green's function (NEGF)
model~\cite{WackerIEEEJSelTopQuant2013,WingeJAP2016}, where we assume
a homogeneous bias drop along the structure allowing for periodic
boundary conditions between the modules. For the device V812, studied
here, we instead find a current peak at a bias of $\sim$39 mV per
module, which is below the NOP, see
Fig.~\ref{FigHomogeneous}(b). Above this point, the current drops with
bias resulting in a region of NDC, which even covers the NOP at 54 mV
per module, where the injector level aligns with the upper laser
level. Thus, without  stimulated emission, the NOP is  not directly
accessible.  However,  the current starts to increase at a higher bias
exhibiting a peak at 65 mV, where the upper laser level aligns with a
further level 5~\footnote{For its most part, level 5 corresponds to
  the second excited state of the lasing double-well.}. We will
demonstrate how this tunnel resonance plays an essential role during
laser ignition as it provides an operation point with positive
differential conductivity exhibiting sufficient optical gain.  For the
self-consistent lasing field [red dashed line in
  Fig.~\ref{FigHomogeneous}(b)], where gain and losses compensate, the
current is strongly increased, resulting in positive differential
conductivity and stable operation around the NOP. This situation is in
complete analogy with the scenario proposed for SLs
\cite{KroemerArxiv2000}.  Comparing Figs.~\ref{FigHomogeneous}(a,b),
we find that the ratio between the lasing-induced current and the
background current is actually higher for the QCL operating in the NDC
region.  For the  bound-to-continuum structure in
Fig.~\ref{FigHomogeneous}(a) further current paths through other
levels prevent NDC. This results in a  parasitic background current
density at the operation point which acts as a source of
dissipation~\cite{ChassagneuxTransTerahertzSciTechnol2012}. For good
diagonal designs such as the structure V812\footnote{The
    ``diagonality'' of a design is related to spatial separation
    between lasing states and is commonly assessed by the oscillator
    strength, here 0.3 in V812 structure. The lower the oscillator
    strength between lasing states, the more ``diagonal'' is the laser
    design.}, the current without lasing is actually low at the NOP,
as a long life-time of the upper laser state is envisaged. Thus the
scenario observed in Fig.~\ref{FigHomogeneous}(b) is actually
beneficial, provided the ignition of the lasing field is guaranteed.

In this manuscript we show, how the ignition occurs in V812 by a
detailed experimental and theoretical study.  For this purpose we
extend the standard simulation schemes for domain formation for
superlattices \cite{WackerPhysRep2002,BonillaRepProgPhys2005} and QCLs
\cite{WienoldJAP2011} by the interplay with the cavity fields, as
taken into account by our microscopic NEGF simulations, see
Appendix~\ref{SecModellingDomain}.  Using this model, we are able to
simulate the ignition of the lasing field in the NDC region. The
current-bias relation exhibits a ``merlon'' at threshold in good
agreement with the experimental data as discussed in
Sec.~\ref{SecDomain}. Studying the time-dependence of the experimental
bias, we observe self-sustained oscillations on the GHz
scale both before and after threshold, see
Sec.~\ref{SecOscillations}. Within our model, these oscillations are
attributed to traveling field domains. Furthermore, we validate the
ignition scenario by the red shift of the lasing spectrum close to
threshold in Sec.~\ref{SecSpectrum}. Details on the simulations and experiments
are given in Appendices~\ref{SecModelling},\ref{SecSetup}, respectively.
Appendices~\ref{AppAddResults},\ref{AppBiasDependence}
provide additional data from simulations and experiments with a different device.
Finally, Appendix~\ref{SecDeconv} addresses the distortion of the oscillation signal 
by the incomplete RF design.
Our
findings clearly demonstrate the ignition scenario in the NDC region,
which results in conventional lasing at the NOP, where the lasing
induced current stabilizes the operation.

\section{Formation of field domains}\label{SecDomain}
\begin{figure*}
\includegraphics[width=\textwidth]{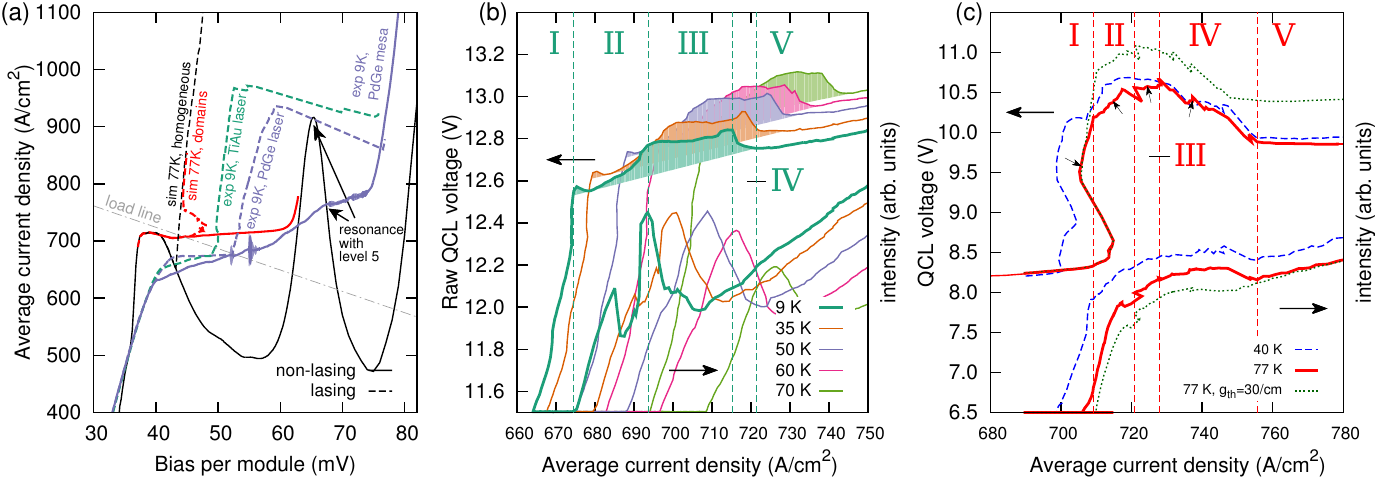}
\caption{(Color online) Current-voltage relations with domain formation: {\bf (a)}
  Quantitative comparison of experimental and simulated
  current-voltage characteristics.  Data taken at 9~K from three
  devices fabricated with different contacts and  waveguide sizes are
  shown. (See Appendix~\ref{SecSetup} for the derivation of
  experimental bias per module.) The three thicker sections in the
  electrical characteristic of the PdGe mesa (purple solid) are
  proportional to the amplitude of voltage self-oscillations, or
  instabilities, as measured with 1 GHz bandwidth oscilloscope. The
  theoretical domain solution (red line) without (solid) and with
  (dashed) the lasing field is compared to the corresponding result
  for a homogeneous bias drop as shown by the black line. The
  dashed-dotted dark grey line denotes the load line in
  Eq.~\eqref{loadline}. {\bf (b)} Zoom in on the experimental
  current-bias relation and lasing intensity  around threshold for
  different temperatures for the TiAu laser. The signification of ``raw'' QCL voltage,
  the vertical axis label, is explained in
  Appendix~\ref{SecSetup}. (The temperatures are ordered from 9 K (left) to 70 K (right) at the bottom of the figure.) {\bf (c)} Simulated results comparing the
  effects of temperature and total losses. In panel (a) and elsewhere
  in the paper, losses of 20 $\textrm{cm}^{-1}$ and a temperature of
  77 K are applied to the simulations. The vertical dashed lines in
  panels (b) and (c) indicate the boundaries between the regions I-V
  at 9~K and 77~K, respectively. The small arrows in (c) mark the
  operation points shown in Fig.~\ref{FigSimOscRegions}.}
\label{FigDomainFormation}
\end{figure*}

%Merlon feature and part a
Fig.~\ref{FigDomainFormation} shows an unusual scenario around
ignition in the  light($L$)-voltage($U_{\textrm{QCL}}$)-current
density($J$) characteristics, where the voltage  has a flat local
maximum (referred to as merlon) around $700~\textrm{A/cm}^2$ as
highlighted by the colored areas in panel (b). Simultaneously, the
light intensity exhibits a peak, while at higher current density
($\gtrsim 800~\textrm{A/cm}^2$) a common linear increase of intensity
and bias is observed. 
The results from our model with domain
formation, see panel (c), agree reasonably well with the experimental
observation in panel (b).

The global scenario leading to the merlon is shown in
Fig.~\ref{FigDomainFormation}(a). The $U_{\textrm{QCL}}-J$ relation is
obtained by operating via a load resistance and a parallel capacitance
with an external bias $U_0$--the output voltage from the pulser. (The
circuit model is further discussed in the
Appendix~\ref{SecModellingDomain}.) If the load line is shifted to the
right of the first current peak around 39 mV per module, a domain
state is established in the QCL, which results in the red solid
horizontal line, also called current plateau. Here the total bias is
distributed on a low- and a high-field domain  [shown in greater
  detail in Fig.~\ref{FigSimOscRegions}(a)]. The high-field domain
exhibits gain (at a frequency slightly higher than the NOP), and the
total gain, as expressed by Eq.~(\ref{EqGainTot}), gets close to the
losses, if the high-field domain becomes large enough. The onset of
lasing changes the local current-field relation, which gives rise to
the pronounced merlon of the dashed red line. For even higher
currents, this domain state merges with the homogeneous state under
lasing (black dashed line) demonstrating that the conventional lasing
around the NOP is stabilized due to the lasing induced current.

%Detailed comparison of Figs 2 b,
The merlon feature occurs for a wide range of temperatures up to 90 K
in our sample, see  Fig.~\ref{FigDomainFormation}(b).  Both in
experiment and simulation, it is shifted to higher currents with
increasing temperature, see Fig.~\ref{FigDomainFormation}(b, c).  In
Fig.~\ref{FigDomainFormation}(c), we also show simulations for an
increased threshold gain $g_\mathrm{th}$, which shifts the bias, but
does not change the essential feature. Such a shift also occurs
between devices with different metallization of the electric contact, see
Fig.~\ref{FigDomainFormation}(a). A similar merlon is also observed
between 89 K and 105 K for the structure V773 of
Ref.~\cite{FathololoumiJAP2013}.

Around the merlon we identify five regions, as indicated in
Figs.~\ref{FigDomainFormation}(b, c).  In region I, $U_{\textrm{QCL}}$
increases  with relatively little variation of current as common for
domain formation, where the increase in bias corresponds to a spatial
increase of the high-field domain. At the experimental threshold, we
observe a kink in $U_{\textrm{QCL}}$ [together with a small spike
  ($\sim$20  mV) in (b) at low temperature]. Subsequently,  a joint
increase of current and bias  marks the region II, the left flank of
the merlon.  Both in experiment and simulation, the lasing intensity
is drastically increasing in this region. This is followed by a region
III, the top of the merlon, where the bias is almost constant and the
calculated lasing intensity shows less variation, while a drop is
observed in the experiment. The latter may be partially related to the
drop in laser frequency occurring here  (see Sec.~\ref{SecSpectrum})
and associated changes in mirror losses. On the right flank of the
merlon, region IV, the bias drops with current. Finally, in region V,
the bias and intensity increase almost linearly with current as common
for lasing operation within a homogeneous field distribution.

%Detailed Comparison of theory and experiment in Fig 2 a
Fig.~\ref{FigDomainFormation}(a) also provides a comparison between the
experimental $U_{\textrm{QCL}}-J$ characteristics of three devices,
processed with either TiAu or PdGe contacts. The PdGe mesa device was
purposely fabricated to frustrate stimulated emission. The
$U_{\textrm{QCL}}-J$ of this device (purple solid line) shows two
pronounced shoulders at average biases of 40 mV and 68 mV per module,
without any detailed structure in between. This confirms our
simulation data (black solid line), which show current peaks close to
these biases, but, without lasing, no current peak at the operation
point of 54 mV per module. Voltage fluctuations, as recorded with a 1
GHz bandwidth oscilloscope, are overlaid on the current density of the
PdGe mesa device in Fig.~\ref{FigDomainFormation}(a). They can be
regarded as a sign for NDC both between the shoulders and after the
shoulder at $\sim$68 mV/module. Laser emission on the low loss
waveguide, i.e. with TiAu contacts, collapses after 55 mV/module,
close to the predicted NOP, a voltage corresponding to the optimum
injection to the upper laser state, which is now accessible
thanks to stimulated emission. This is in contrast with the
higher loss waveguide device, i.e. with PdGe contacts, where laser
emission switches off later, above 60 mV/module.

The simulated increase of current above 75 mV per module is due to
tunneling from the lower lasing level to level 5 (aligned at $\sim$83
mV per module) and subsequent leakage to continuum.
As the current in the continuum is not well covered in our model,
the simulations
underestimate the current here.  Overall, the simulated current
densities and biases agree well with the experimental data.

\section{Oscillating field domains}\label{SecOscillations}

\begin{figure*}
  \includegraphics[width=\textwidth]{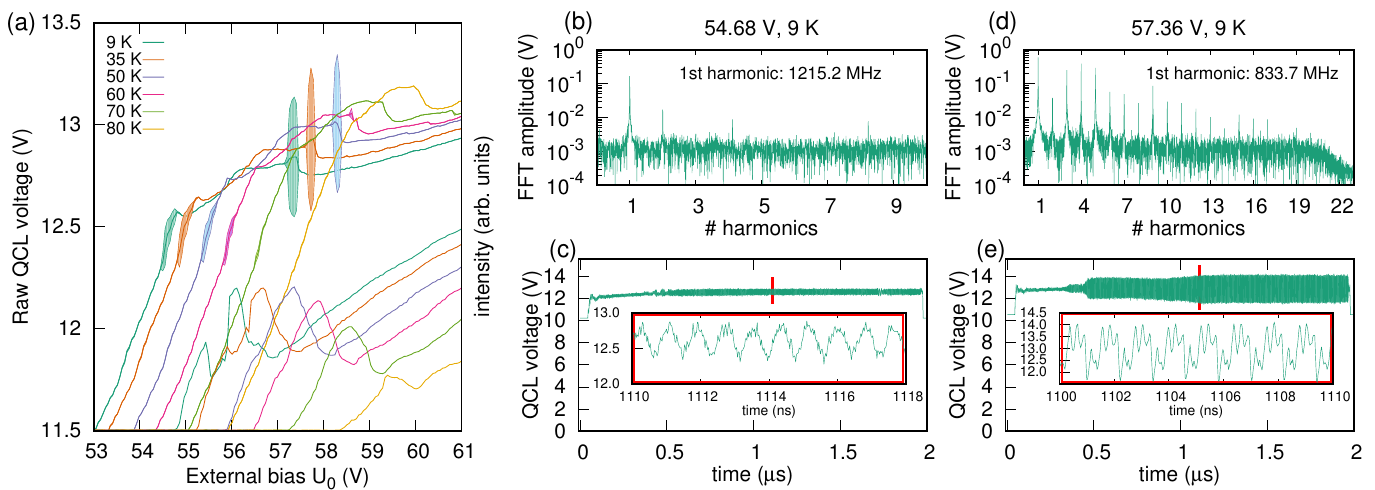}
  \caption{(Color online) Oscillatory behavior detected experimentally: Panel {\bf
      (a)} shows the same data as presented in
    Fig.~\ref{FigDomainFormation}(b) but plotted against external bias
    $U_0$. (The temperatures are ordered from 9 K (left) to 80 K (right) at the bottom of 
    the figure.) The shaded regions indicate voltage oscillations monitored
    by an oscilloscope with 1 GHz bandwidth, which are observed in two
    different ranges of bias for temperatures below 70 K. For 9 K, we
    show the QCL voltage, as measured with a 15 GHz bandwidth
    real-time oscilloscope, during the 2-$\mu\mathrm{s}$ pulse at
    $U_0=54.68~\mathrm{V}$ and $57.36~\mathrm{V}$ in panels  {\bf (b,
      c)} and {\bf (d, e)}, respectively, which are representative for
    the two different regions of oscillation. The amplitude spectrum
    shown in panels (b, d) are taken for the time interval
    $1.08~\mu\mathrm{s}\leq~t~\leq~1.28~\mu\mathrm{s}$ of the pulses
    displayed in panels (c, e). In panel (d) up to 20 harmonics can be
    observed, hence demonstrating the sharp and small features in the
    time-resolved voltage in panel (e) are highly periodic. Due to
    incomplete RF design, details of the oscillations on the sub-ns
    timescale in panels (c, e)  may be improperly recorded, see
    Appendices \ref{SecDeconv} and \ref{SecSetup}.}
\label{FigExpOsc}
\end{figure*}

%Experimental oscillations
Fig.~\ref{FigExpOsc} shows self-sustained bias oscillations measured
experimentally. As exemplified in panels (c) and (e), they occur both
below and above threshold. The shaded regions in panel (a) show the
extend of the oscillatory behavior as monitored with the 1 GHz
bandwidth oscilloscope, see Appendix~\ref{SecSetup}.  With a high
bandwidth real-time oscilloscope, we observe GHz oscillations up to a
temperature of 78 K in parts of region III and the transition to
IV. Here, we measure frequencies in the range of $0.8-0.9$ GHz, which
are decreasing with bias for each temperature. In region I, the less
intense GHz oscillations are detected up to 68~K. The frequencies are
around 1.2 GHz and increase up to slightly 1.3 GHz above
threshold. The laser ignites while the voltage self-oscillates,
nevertheless these oscillations are quickly damped as the bias
increases above threshold (beginning of region II), see also Fig.~ \ref{FigFreqVsTime1}
in Appendix~\ref{AppBiasDependence} for the oscillations around
threshold. The measurement of the bias dependence of the
oscillation frequency is described in Appendix~\ref{AppBiasDependence}.

The occurrence of oscillations with GHz frequency can be understood
on the basis of traveling field domains as known from Gunn
diodes~\cite{ShawBook1992} and
superlattices~\cite{WackerPhysRep2002,BonillaRepProgPhys2005}. For the
low doping density, the NDC region cannot be overcome by charge
accumulation in a single module
\cite{WackerPhysRep2002,WackerPRB1997a}.  Based on the current and
electron density, we estimate an average drift velocity of 6.6 km/s,
with which these excess carriers between the domains travel through
the structure.
As a new accumulation layer forms only after the preceding
one has traversed a large part of the device with a length of 10 $\mu$m,
we get a typical oscillation period of 1 ns.
    
\begin{figure}
\centering
\includegraphics[width=0.9\columnwidth]{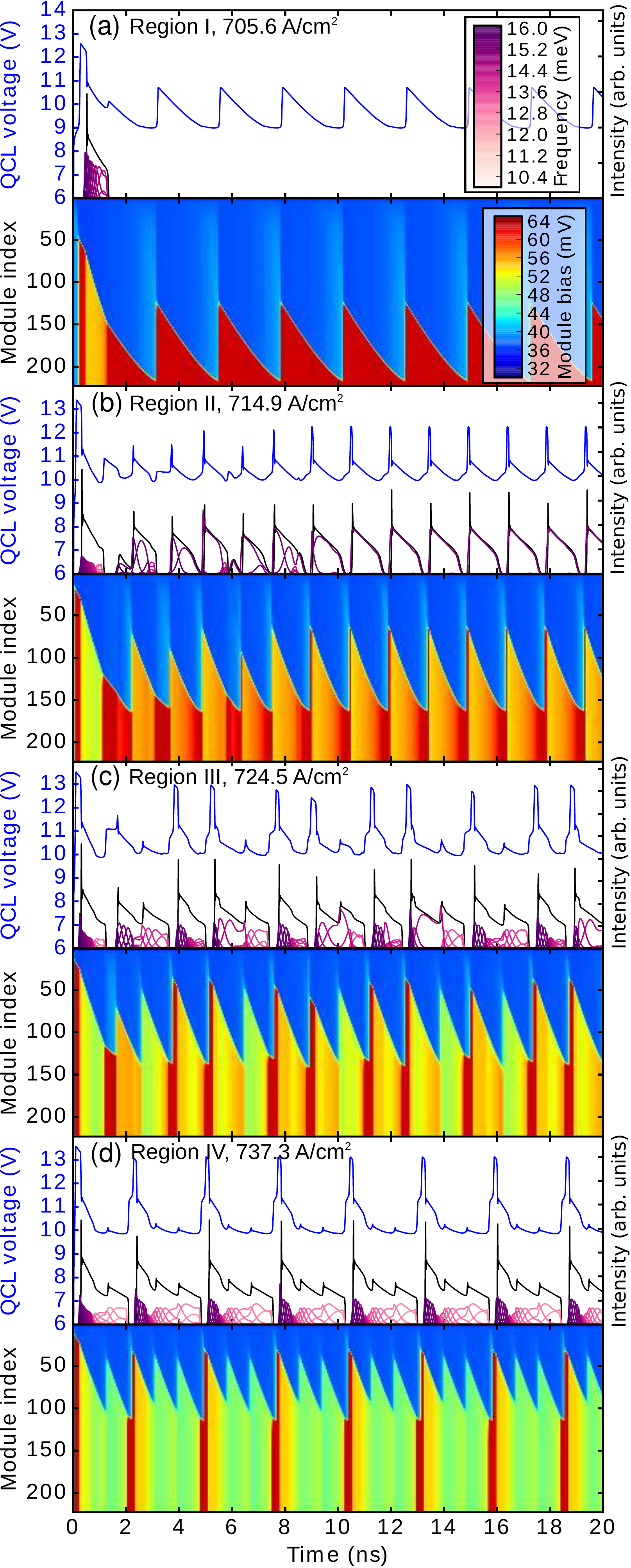}
\caption{(Color online) Simulated oscillations for four different operation points
    as indicated in
    Fig.~\ref{FigDomainFormation}(c). For each panel, the
  space and time dependence of the electric field is displayed in color
  scale. The bottom module ($m=222$) is next to the positive contact.
  The upper part of each panel shows the bias along the QCL
  structure (blue) and the total lasing intensity (black), which is also resolved for the individual modes with frequency marked by color. Data for
  further operation points and Fourier transforms of the
  voltage signals are displayed in Appendix~\ref{AppAddResults}.
}
\label{FigSimOscRegions}
\end{figure}

%Details on simulated domains
Such oscillations based on traveling field domains are found in our
simulations as shown in Fig.~\ref{FigSimOscRegions} for a selected
set of operation points in the current-bias relation (extensive data
can be found in Appendix~\ref{AppAddResults}. Here we find
different oscillation modes in the different regions of the
current-bias characteristics, labeled I-IV in
Fig.~\ref{FigSimOscRegions}. The reason is that the local
current-field relation is strongly modified by the lasing field.

%Region I
In region I, the lasing field does not play a role
  and the local current-field
relation [as given by the black line in Fig.~\ref{FigHomogeneous}(b)]
is unaltered.  Within the domains, the condition of a constant current
density implies the  bias drop per module of $\approx 38$ mV and
$\approx 62$ mV, in the low- and high-field domain, respectively.
Between both domains, there is a layer of electron accumulation
[e.g. around module 150 at 6 ns in Fig.~\ref{FigSimOscRegions}(a)], which travels towards the positive
contact. From time-resolved electron density plots we find that the
accumulation layer spans over several modules ($\sim$3-6). The
movement of the accumulation front provides a drop in bias and, following
the load line, an increase in current. Eventually, the current reaches
the first peak in the local current-field relation, where the
low-field domain becomes unstable and a new domain boundary forms
(e.g. around module 130 at 8 ns). These oscillations exhibit
frequencies of 0.4-0.6 GHz [as can be seen in Figures~ \ref{fig:sim_osc_fft_0},
 \ref{fig:sim_osc_fft_1} of Appendix~\ref{AppAddResults}], increasing
with external bias $U_0$ in accordance with measurements.

%Region II
In region II, the lasing field is strong enough to alter the local
current-relations resulting in the surge for the average current, see
Fig.~\ref{FigDomainFormation}. Lasing sets in just after the forming
of the high-field domain [e.g. at 12 ns in
  Fig.~\ref{FigSimOscRegions}(b)]. The increased current results in a
drop of field in the high-field domain, as conduction plus
displacement current has to match the current in the low-field
domain. In the following, the accumulation front travels, reducing the
extend of the high-field domain and the total gain drops, so that
lasing operation stops. While the accumulation front travels further,
the low-field domain reaches the first current peak and a new domain
boundary forms similar to region I.

%Region III
Upon increasing the external bias to region III, the new instability
may occur, while lasing is still active. Then the new high-field
domain arises at a lower field around 50 mV per module [e.g. at 10 ns
  in Fig.~\ref{FigSimOscRegions}(c)]. However, starting from this
state, lasing always stops before the next domain forms and we observe
an irregular (possibly chaotic, but we did not prove this) pattern in
Fig.~\ref{FigSimOscRegions}(c) or different types of locking between
both scenarios, see Fig.~\ref{fig:sim_osc_fft_2} in Appendix~\ref{AppAddResults}.

%Region IV
In region IV, see Fig.~\ref{FigSimOscRegions}(d), lasing prevails most
of the time and at least half of the high-field domains are formed
close to the NOP with fields around 50 mV per module. This provides
the drop in average bias as characteristic for this region. While new
domains form about once per ns, the fundamental oscillation frequency
is 0.37 GHz  in Fig.~\ref{FigSimOscRegions}(d) as only every third
time the lasing operation stops.

%Comparison Exp Sim oscillation
Comparing with the experimental oscillations, the shape of the bias
signal in region I is triangular both in experiment,
Fig.~\ref{FigExpOsc}(c), and simulations,
Fig.~\ref{FigSimOscRegions}(a). The bias signal in regions III/IV
resembles more a switching between plateaus with bias values separated
by 1 V.  Here the experimental data, Fig.~\ref{FigExpOsc}(e), do not
show the thin bias spikes visible in Figs.~\ref{FigSimOscRegions}(c,
d). Instead, a strong ringing-like feature with a pseudo-period of
$0.2-0.3$ ns is observed, which is attributed to lacking RF design (see 
Appendix~\ref{SecSetup}). An attempt to filter out ringing effects
  has been tested and some deconvolution results are reported in
  Appendix~\ref{SecDeconv}. The simulated frequency of the
oscillations in region I ($\sim$0.5 GHz) is lower than the
experimental results ($\sim$1.2 GHz), while in region III the
experimental frequency well matches the mean rate of domain formations
(both $\sim$0.8 GHz). In region III, we also detected a subharmonic regime
 at a specific bias range, see Fig.~\ref{FigExpSubharmonics}  
 in Appendix~\ref{AppBiasDependence}  and Fig.~\ref{FigDeconvExample}(b) 
 in Appendix~\ref{SecDeconv}, which might indicate different coexisting domain
ignition processes as observed in the simulations in regions III and
IV. The main discrepancy is that  the simulations provide oscillations
throughout all the regions I-IV, while experimentally they are only
clearly detected within parts of the regions I, III, and IV (the
latter is small in the experiment), as shown in
Fig.~\ref{FigExpOsc}(a). However, we experimentally observed strongly
enhanced noise on voltage, light intensity, and phase on the lock-in
amplifier used during THz detection in regions II and III, which we
could not fully explain yet. In this context, we note that the
stability of domains depends both on the details of the electron
distribution in the transition region between the domains, the nature
of the injecting contact, and the external circuit (for instance the
50 Ohm coaxial cable was not considered in simulations). For these
issues, we did rather simple approximations, which are difficult to
overcome as no consistent treatment has been proven valid so
far. Furthermore, pinning of the accumulation layer may occur if
individual modules differ due to growth imperfections. These effects
could also cause the differences between the measured and simulated
oscillation frequencies.

\section{Lasing Spectra}\label{SecSpectrum}

\begin{figure}
\includegraphics[width=\columnwidth]{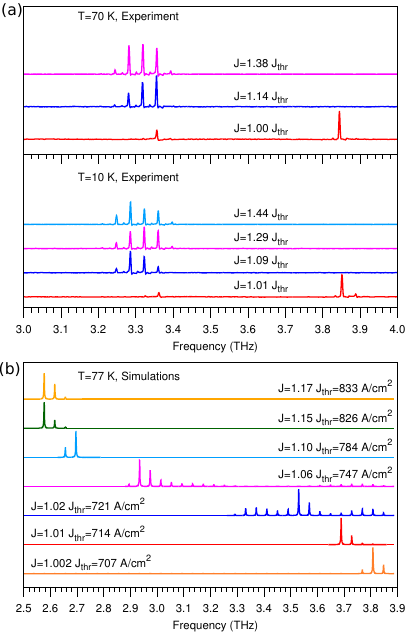}
\caption{(Color online) Lasing spectra demonstrating a significant red-shift
  after threshold:  {\bf (a)} Experimental spectra for the
  sample V812 at different temperatures and currents. Close to threshold, a
  high frequency signal can be seen at 3.84 THz, while lower
  frequencies around 3.3 THz are observed for higher current. {\bf
    (b)} Simulated spectra at different current densities plotted with
  a linewidth of 0.2 GHz. Lasing starts in a high-field domain, see,
  e.g., Fig.~\ref{FigSimOscRegions}(b), and is shifted towards the
  lower frequencies as current increases.}
\label{FigSpectra}
\end{figure}

For the ignition scenario from a domain state discussed here, the
high-field domain exhibits a higher bias drop per module than the
final lasing state.  Due to the Stark shift, which is substantial for a
diagonal design as V812, we thus expect a higher lasing frequency at
ignition. Fig.~\ref{FigSpectra} shows that this is indeed the
case. Just after threshold, the lasing frequency is around  3.8 THz in
the experiment and slightly lower in the simulation. With increasing
current a drop to 3.3 THz (experiment) or 2.6 THz (simulations) is
observed, where the frequencies remain constant with further increase
of current.  The discrepancy in frequency for operation in region V
is related to the lower threshold bias in the simulations.
Fig.~\ref{FigDomainFormation}(a) shows, that the lasing around 800
A/cm$^2$ occurs at 45 mV per module in the simulations and at 50 mV
per module in the experiment. Thus the difference in frequency
corresponds to the Stark shift. E.g., for an operating bias of 52 mV
per module, the simulations show a gain peak at 3.1 THz (see also 
Fig.~\ref{FigHomsimulation} in the Appendix~\ref{SecFitting})
in much better agreement with the
experimentally observed lasing. A similar spectral red-shift after threshold was also observed in the more diagonal structure V773 of Ref.~\cite{FathololoumiJAP2013}.

\section{Conclusion and Outlook}
The onset of lasing from a state of oscillating electric field domains was
demonstrated for THz QCLs. Comparing experimental and theoretical
results, this feature is clearly reflected in the shape of the
current-bias relation, characteristic self-sustained oscillations
around threshold, and the red shift in the optical spectra just after
threshold.

This ignition type demonstrates that there is no need for positive
differential conductivity for the NOP at threshold, which is also
visible in other highly performing QCLs~\cite{ChanAPL2013,
  FathololoumiJAP2013}. Instead, significant gain at a resonance at
higher field is sufficient. In this case domain formation in the NDC
region allows a part of the device to operate at this auxiliary
operation point. After ignition of the lasing field, the stimulated
transitions provide sufficient current at the NOP to break the
NDC. Thus a conventional requirement for laser design is lifted, which
allows for more flexibility in the design of QCLs at the price of
noise-like behavior at threshold. In particular, such a new design
could reduce the background current and provide more effective power
conversion to the lasing field. In order to realize this scenario,
during the design stage of highly diagonal structures particular
attention should be paid to the gain at the positive differential
resistance region next above the NOP.

Furthermore, we found complex oscillation patterns due to the
interplay between the lasing field and the running field domains. This
establishes a further degree of freedom compared to the related
superlattices with their rich spectra of nonlinear and chaotic
behavior\cite{AmannPRL2003,FromholdNature2004,HramovPRL2014}. In this
context the time-resolved measurement of the lasing activity with a
fast THz detector~\cite{ScheuringTransTerahertzSciTechnol2013} could
provide additional relevant data.

\begin{acknowledgments}
  The authors would like to thank Dr. Saeed Fathololoumi for
the FTIR measurements and Dr. Seyed Ghasem Razavipour
for fruitful discussions and for providing unpublished data on
wafer V773 collected in the lab of Prof. Dayan Ban, Waterloo University. Dr. Sergei Studenikin and Nick Donato from
Tektronix have kindly lend us the 15 GHz and 2 GHz bandwidth oscilloscope, respectively. Financial support from the Swedish Science
Council (grant 2017-04287) is gratefully acknowledged.
\end{acknowledgments}

\appendix

%%%%%%%%%%%%%%%%%%%%%%%%%%%%%%%%%%%%%%%%%%%%%%%%%%%%%%%%%%%%%%%%%%%%%%%
\section{Simulation procedure} \label{SecModelling}
\subsection{NEGF simulations for homogeneous field distributions}
Our calculations are based on the NEGF model
\cite{WackerIEEEJSelTopQuant2013,WingeJAP2016} providing us with
the current density $J(F_\mathrm{dc},F_\mathrm{ac},\omega_0)$ and gain
$G(F_\mathrm{dc},F_\mathrm{ac},\omega_0)$ which are nonlinear
functions of the electric field $F(t) = F_\mathrm{dc} + F_{\rm ac}
\cos(\omega_0 t)$. Here $F_\mathrm{dc}$
is the field due to an applied bias and $F_{\rm ac}$ is the electrical
component of the lasing field in the waveguide with frequency
$\omega_0$. As common for most simulation schemes \cite{JirauschekApplPhysRev2014}, these calculations assume a homogeneous bias drop along the QCL structure.
To calculate the current under operation we increase the ac field strength
at each bias point until gain is saturated by the level of the 
losses $g_{\rm th}$. In addition, the peak gain frequency is 
updated with the ac field strength to track intensity dependent 
effects on the laser transition. This procedure provides the data presented
in Fig.~\ref{FigHomogeneous}(a,b). 
The temperature used in the
simulations defines the thermal occupations of the phonon modes. Due
to the excitation of the mostly relevant optical phonons on short time
scales~\cite{VitielloAPL2012,ShiJAP2014} this temperature should be
higher than the experimental heat sink temperature. As input we use the
nominal sample parameters together with an exponential interface roughness
model\cite{FranckieOE2015} with 0.2 nm height and 10 nm lateral correlation length. All these model details agree with \cite{WingeJAP2016}, where results for
a large number of devices are shown. 
Note, that we
define the bias $U_{\rm QCL}$, electric field $F$ and electric current(density)
$I(J)$ with an additional minus sign throughout the paper to
compensate for the negative electron charge $-e$.

\begin{figure}
  \includegraphics[scale=0.235]{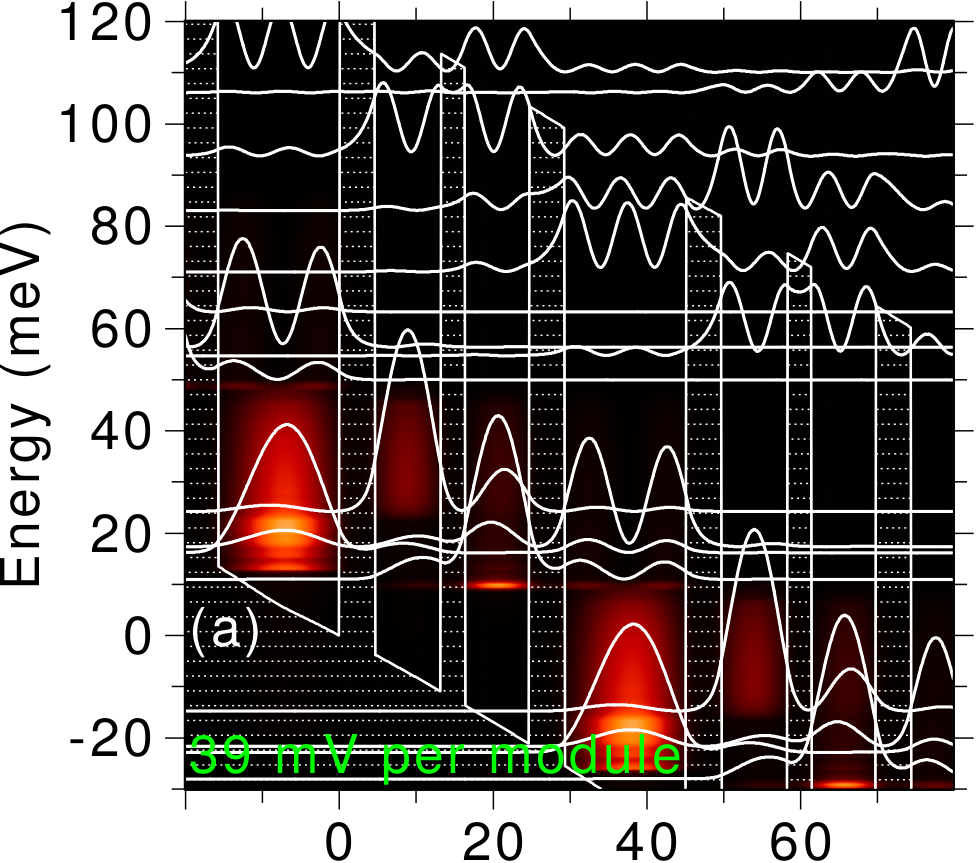}
  \includegraphics[scale=0.235]{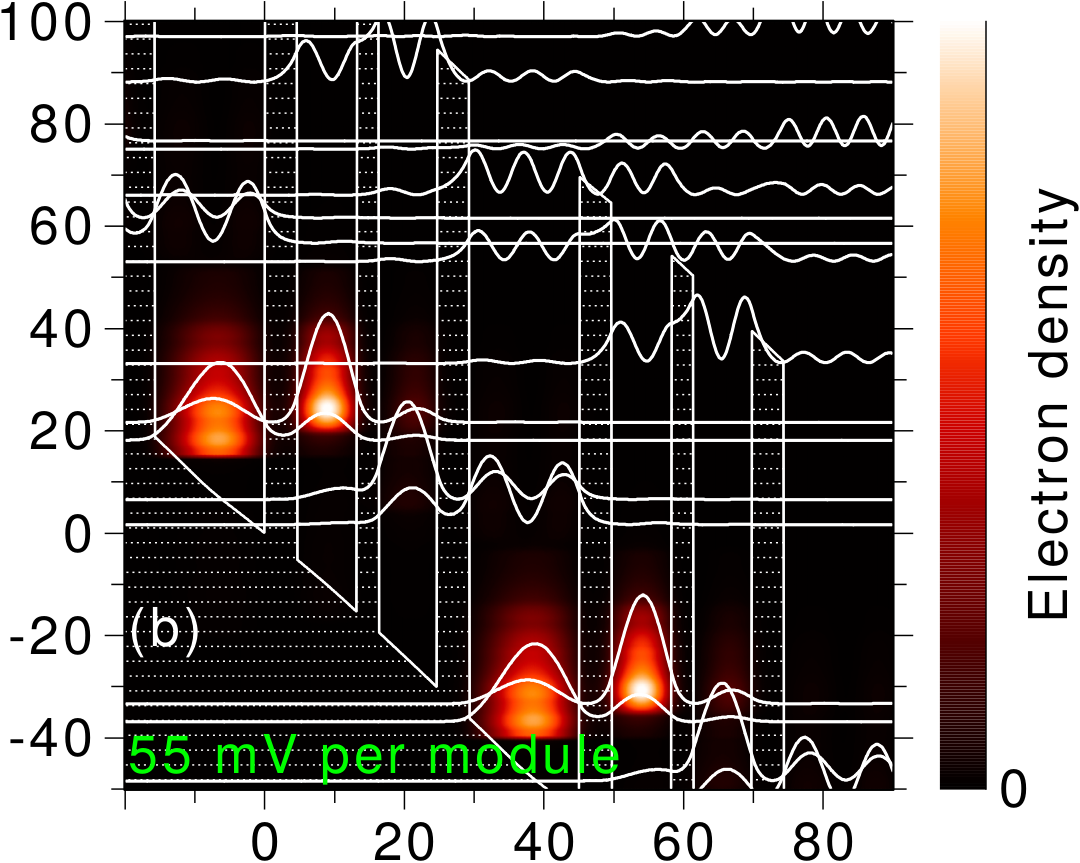}\\
  \includegraphics[scale=0.235]{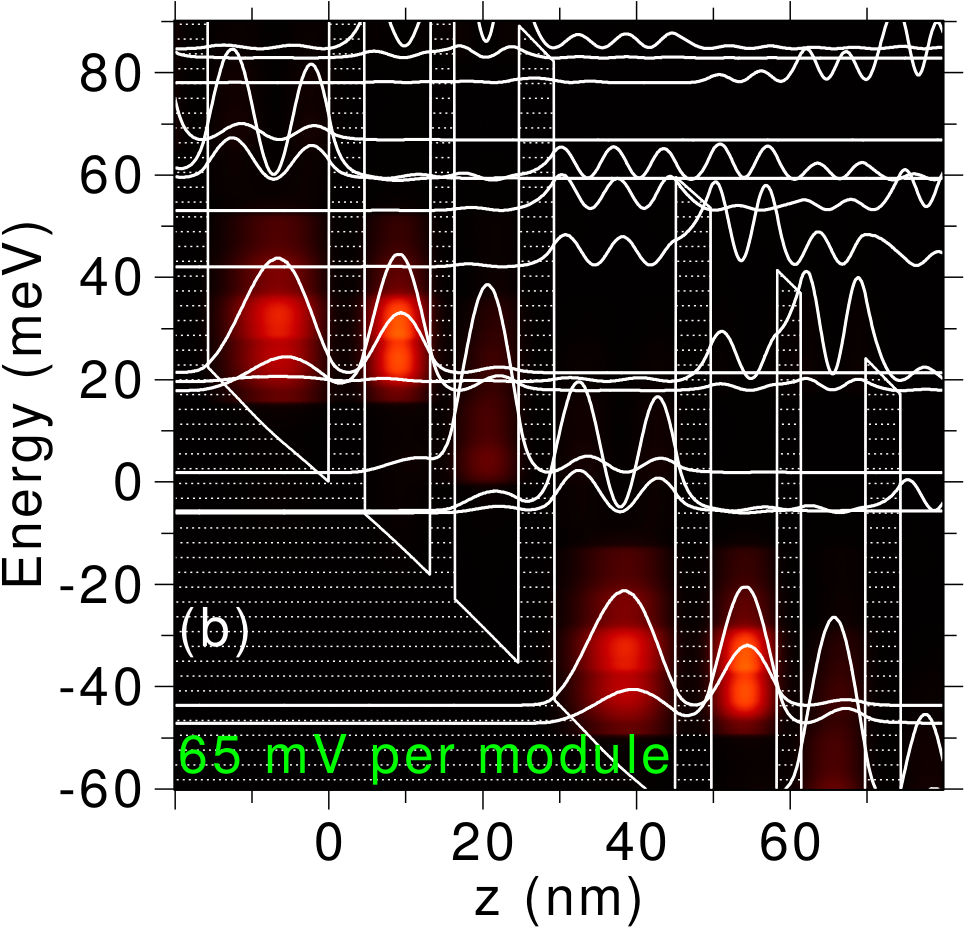}
  \includegraphics[scale=0.235]{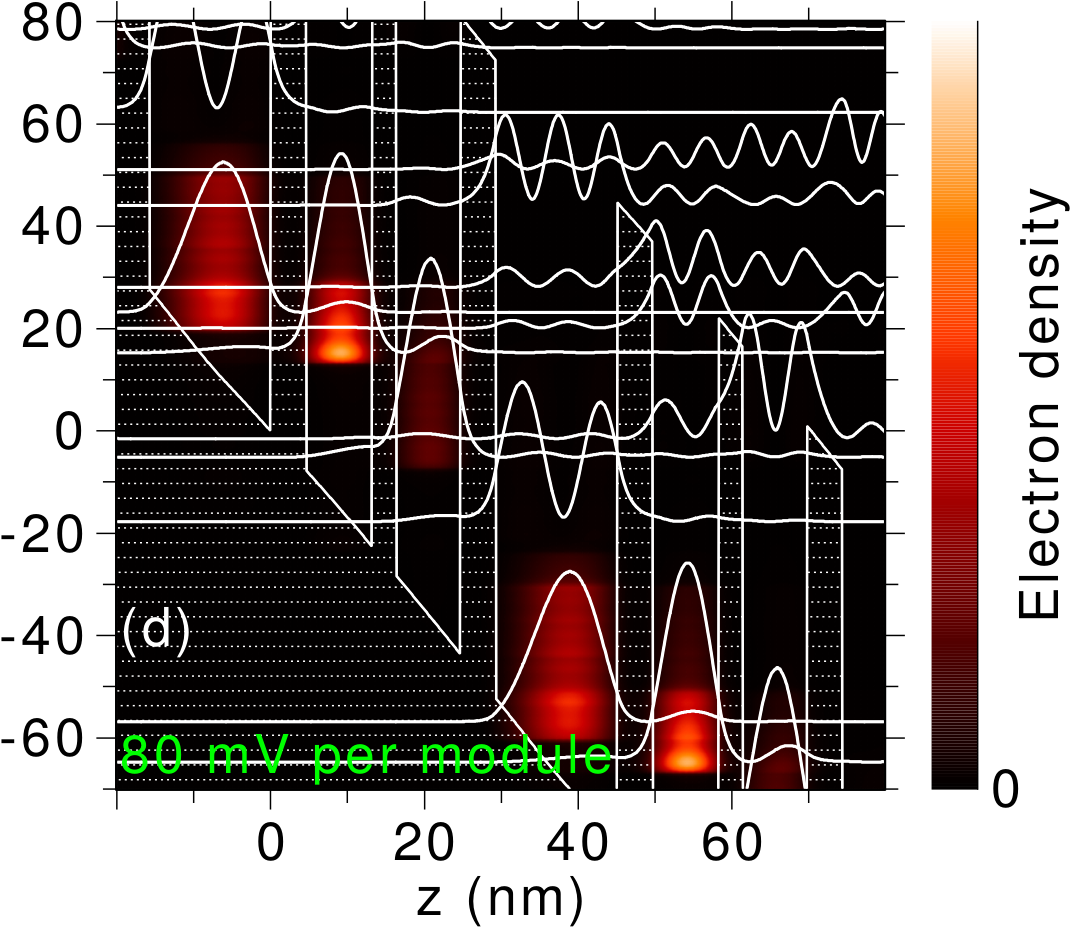}
   \caption{(Color online) Electron density calculated by the NEGF approach for different
biases per module together with the Wannier-Stark levels.}
   \label{FigDensityplots}
\end{figure}

In Fig.~\ref{FigDensityplots} we show the calculated electron
densities together with the Wannier-Stark levels for different biases,
in order to study the relevant alignments. For a bias of 39 mV per
module (panel a) the injector states aligns with the lower laser level
and the extractor state allowing for a strong current through the
entire module. This is just the first current peak in
Fig.~\ref{FigHomogeneous}(b). The resonance at 65 mV is shown in panel
(c) and the increasing current around 80 mV [see
  Fig.~\ref{FigDomainFormation}(a)] is attributed to tunneling to
continuum states in panel (d).  Here, we used 9 states per module and
next-nearest neighbors in the simulations. For biases below 70 mV per
module, we found that 7 states per module were sufficient, which
allows to reduce the numerical effort.

In order to simulate domain formation with the interplay of an
intra-cavity lasing field, as shown in Fig.~\ref{FigDomainFormation},
the gain needs to be estimated at frequencies $\omega$, which may
differ from $\omega_0$ of the lasing field.  A common situation in
this study is that lasing sets in in the high field domain, allowing
for a build up in intensity at the frequency favored at that field
strength.  As intensity increases, the local current-field relation
changes to a degree, where lasing at other frequencies is more
favorable compared to the original one.  This poses the question: How
does the established laser field promote the onset of lasing at other
frequencies? To answer this, multidimensional fitting of the simulated
NEGF data is not enough, as  we need to estimate the gain at
frequencies other than the lasing frequency. Our approach is to use
fits for the current density
$J^\textrm{fit}(F_\mathrm{dc},F_\mathrm{ac},\omega_0)$ and gain
$G^\textrm{fit}(F_\mathrm{dc},F_\mathrm{ac},\omega_0,\omega)$ which
are based on a simple physical model (see Appendix~\ref{SecFitting})
and where the parameters are determined by comparing with the full
NEGF calculations.

\subsection{Domain formation and external circuit}\label{SecModellingDomain}

\begin{figure}
\includegraphics[width=\columnwidth]{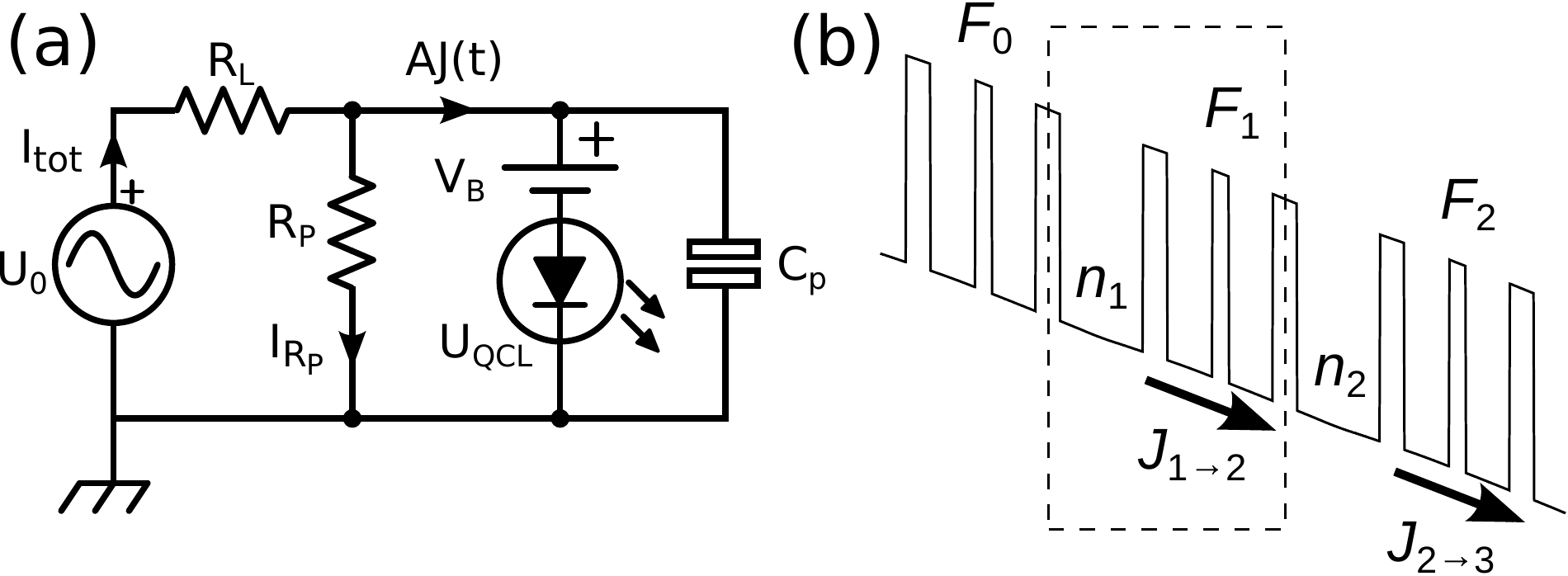}
\caption{{\bf (a)} Circuit including a parasitic capacitance $C_p$ in
  parallel to the QCL as well as a voltage probe of resistance $R_p$,
  a load resistance $R_L$ and a Schottky voltage barrier $V_B$ as well
  as the pulser at an external bias of $U_0$. In addition to this
  model, a 50 Ohm coaxial cable connects the pulsed bias source to the
  load which is neglected here. The voltage probe resistance $R_p$
  includes the 50 Ohm input impedance of the oscilloscope. {\bf (b)}
  Numbering of electron densities $n_m$, fields $F_m$, and current
  densities $J_{m\to m+1}$.} 
\label{FigCircuitNumbering}
\end{figure}

In this section, we describe the simulation of the extended QCL
structure with $N=222$ modules within the circuit shown in
Fig.~\ref{FigCircuitNumbering}(a).
In particular, we drop the
assumption of a periodic voltage drop along the QCL structure, which
lifts the requirement for charge neutrality in each module of the QCL.
Instead, the electron density $n_m$ in module $m$ may differ from the
background doping $n_D$ (both in units 1/area). Here we assume that
the additional charge of each module is essentially located in the
injector well [the widest well of the structure, see
  Fig.~\ref{FigHomogeneous}(c)], and that the current density
$J_{m\to m+1}$ between the injector wells $m$ and $m+1$ essentially
depends on the bias drop $F_md$ over module $m$. The numbering is
illustrated in  Fig.~\ref{FigCircuitNumbering}(b).  Following
Ref.~\cite{WackerJAP1995}, Ampere's law and Poisson's equation provide
\begin{align} \nonumber
\epsilon_r\epsilon_0 \frac{\d F_m}{\d t } =& J(t) - J_{m\to m+1}
\\ \label{diff1} &- \frac{C_{p}} {C_s+C_p} \left( J(t) - \frac{1}{N+1}
\sum_{i=0}^N J_{i\to i+1} \right) \\ \label{diff2} n_m =& n_D +
\frac{\epsilon_r \epsilon_0}{e} (F_m-F_{m-1}) 
\end{align}
where  $J(t)$ is the current density flowing into the QCL device and
$C_s=A\epsilon_r\epsilon_0/d(N+1)$ is the intrinsic QCL capacitance
assuming an average  relative dielectric permittivity $\epsilon_r=12.9$ for
the active region with the cross section $A=0.15~{\rm mm}^2$ 
 and a module thickness
$d=45.07$ nm. We use the parasitic capacitance $C_p=C_s/4$ and checked
explicitly that larger values like $C_p=4C_s$ hardly change the
results. Operating via a load resistor $R_L$, the circuit in
Fig.~\ref{FigCircuitNumbering}(a) provides
\begin{align}
  \label{loadline}
     AJ(t) &= \frac{U_0}{R_L} - (U_{\rm QCL}(t) +
    V_B) \left( \frac{1}{R_p} + \frac{1}{R_L} \right)
    \\ &\textrm{with } \quad U_{\rm QCL}(t)=\sum_{m=0}^N F_m(t)d\, ,
\end{align}
where $V_B$ is the Schottky barrier and $R_p$ is the voltage probe
resistance. Throughout the paper we use $R_L=41.2$ Ohm, $R_p=1050.4$
Ohm, and $V_B=0.8~V$ in our simulations.  The load curve of
Eq.~\eqref{loadline} does not consider the effect of the residual
resistance of the ground electrode on Si carrier, as addressed in Appendix~\ref{SecSetup}.  Determining the
current density  $J_{m\to m+1}$ within a module is far from trivial,
as we have to take into account a non-periodic situation and carrier
densities differing from the doping. Following the common
approximation for superlattices
\cite{WackerPhysRep2002,BonillaRepProgPhys2005} and QCLs \cite{WienoldJAP2011}, we use
\begin{equation}
  J_{m \to m+1}= J^\textrm{fit}(F_m,F_{ac},\omega_0)
  \frac{n_{m}-n_{m+1} \e^{-eF_md/k_{\rm B}T} }{ n_D-
    n_D\e^{-eF_md/k_{\rm B}T} }\, .
    \label{EqJmmplus1}
\end{equation}
This expression is only defined for $1\le m< N$, as a real
injector well is required on both sides.  In order to close the
equation system, we further need the cases  $m=0$ and $m=N$. Here we
assume Ohmic boundary conditions
\begin{equation}
J_{0\to1} = \sigma_l F_0 \quad \mbox{and }
J_{N\to{N+1}} = \sigma_l F_N \frac{n_N}{n_D}
 \label{EqDom6}
\end{equation}
with the conductivity $\sigma_l=0.15\textrm{ A/Vcm}$, which is chosen to be
higher
than the average conductivity in the QCL structure. This modeling follows essentially \cite{WienoldJAP2011}, where domain formation in QCLs was studied. Going beyond this work, we also consider the coupling to the lasing field here.

The waveguide modes constitute a global coupling between the modules.
Each mode $i$ is associated with a constant frequency $\omega^i_0$, an
electric field strength $F^i_{ac}$, and its average photon number
$N^i_\mathrm{ph}\propto (F_\mathrm{ac}^i)^{2}$.  Similarly, the gain
in the waveguide is obtained by the average over all modules, where we
assume that the gain is proportional to the electron density in the
respective module:
\begin{equation}
 \bar{G}(\omega,\{N^i_\mathrm{ph}\}) = \frac{1}{N}\sum_{m=1}^N \frac{n_m}{n_D}
  G^\textrm{fit}(F_m,\{N^i_\mathrm{ph}\},\{\omega_0^i\},\omega)\, .
  \label{EqGainTot}
\end{equation}
Here the gain in each module is subject to saturation
from all cavity modes.
The gain recovery time is given by the time-scales for scattering
and tunneling in the modules, which are typically $\lesssim 1$ps. This is
shorter than the photon lifetime of 6 ps (for the gain threshold of $20\, \textrm{cm}^{-1}$), so that we can assume,
that the gain is an instantaneous function of the intensity. For the photon
numbers we thus have 
\begin{equation}
  \frac{\d N_\mathrm{ph}^i(t)}{\d t} = 
  \left(  \bar{G}(\omega,\{N^i_\mathrm{ph}\}) - g_{\rm th}
\right)  \frac{c}{n_g} N_\mathrm{ph}^i(t) 
+ \sum_m \frac{A n^{\rm ULS}_m}{\tau^i_{\rm sp}} 
\label{EqPhotonBalance}
\end{equation}
for all cavity modes $i$, where $n_g=3.6$ is the group refractive
index (assumed to be constant here), and $g_{\rm th}=20\, \textrm{cm}^{-1}$ is given
by the waveguide and mirror losses. A similar approach to model the
photon density in a THz QCL was reported in Ref.~\cite{AgnewAPL2015}.

The last term in Eq.~(\ref{EqPhotonBalance}) describes  the
spontaneous emission of light from electrons in the upper laser
level into the lasing mode $i$, which  is
crucial for ignition. Assuming a
plane wave in our metal-metal waveguide, we can use the standard
textbook expression for the spontaneous emission rate [see, e.g., Eq.~(8.3-6) of \cite{YarivBook1989}]
\begin{align}
  \frac{1}{\tau^i_{\rm sp}}
  =\frac{\pi\e^2\omega_i}{\epsilon_r\epsilon_0 AL}
  |z_{lu}|^2\mathcal{L}(\hbar \omega_i - E_u+E_l)\, ,
\end{align}
where  $z_{ul}$ is the matrix element between the upper and lower
laser state, $L=10\, \mu$m is the thickness of the waveguide, and
$\mathcal{L}$ is the Lorentzian from Eq.~(\ref{EqLorentzian})
replacing the delta function. For different operation points, we find
typical values $\tau^i_{\rm sp}\approx 3-6 $ ms. For simplicity we
choose $\tau^i_{\rm sp}\approx 3$ ms for all simulations.  This value
is appropriate for the high-field domain with $eF_\textrm{dc}d\approx
62$ meV, where lasing typically sets in. (Here $\omega_i$ is large and
$\Gamma$ is small, see Fig.~\ref{FigHomsimulation}). We checked
explicitly, that increasing $\tau^i_{\rm sp}$ by a factor of two
hardly changed the results.

\subsection{Effective model of a single module}\label{SecFitting}

In order to fit the current and the gain, we use expressions resulting
from a simple two-level model, onto which the NEGF results are mapped.
We consider an injecting current $J_{\rm inj}$ into level 2,
scattering lifetimes $\tau_1$, $\tau_2$ for level 1 and 2,
respectively, and $\tau_{21}$ for scattering from 2 to 1.  A laser
field closely resonant to the energy difference $E_{21}$ induces a
transition rate, which, according to Fermi's golden rule
\cite{SakuraiBook1993}, is proportional to the square of the
oscillating field:
\begin{align}
 W_{2\to 1}&=\gamma_{12}(\omega)F^{\,2}_{\rm ac} \\
 \mbox{with } \gamma_{12}(\omega)&= \frac{2\pi}{\hbar} \left|\frac{ez_{21}}{2}\right|^2  
 \mathcal{L}(E_{21}-\hbar\omega)\nonumber
\end{align}
where the energy-conserving $\delta$-function is replaced with the more
realistic Lorentzian 
\begin{equation}
 \mathcal{L}(E)=\frac{1}{2\pi}\frac{\Gamma}{E^2 + \Gamma^2/4}  
 \label{EqLorentzian}
 \end{equation}
 Under irradiation with $\omega_0$ we can thus establish the rate
equations, following Ref.~\cite{FaistBook2013}, 
\begin{align} \nonumber
  \frac{\d n_2}{\d t} &= \frac{J_{\rm inj}}{e} -\frac{n_2}{\tau_{2} } - \gamma_{12}(\omega_0)F_{\rm ac}^{\,2}(n_2-n_1) \\ \nonumber
  \frac{\d n_1}{\d t} &= \frac{n_2}{\tau_{21}}- \frac{n_1-n_{1}^{\rm th}}{\tau_1} + \gamma_{12}(\omega_0)F^{\,2}_{\rm ac}(n_2-n_1)  
\end{align}
where the
second term of the equation for $n_1$ makes sure that the population
reaches its thermal equilibrium value $n_1^{\rm th}$ in the absence of
the injecting current and ac field.  For the steady state, we obtain the
inversion $\Delta n_{21}=n_2-n_1$
\begin{align} \nonumber
  \Delta n_{21} =& \tau_2 \frac{\tau_{21}-\tau_1}{\tau_{21}}\frac{J_{\rm inj}}{e}- n_1^{\rm th} \\ \label{eq:inv_twolevel} &- \left(\tau_1+  \tau_2 \frac{\tau_{21}-\tau_1}{\tau_{21}}\right) \gamma_{12}(\omega_0)F^{\,2}_{\rm ac}\Delta n_{21}
\end{align}
where right hand side of the first line corresponds to the inversion
$\Delta n^0_{21}$ for vanishing ac field strength. Thus
\begin{equation}
  \Delta n_{21}=\frac{\Delta n^0_{21}}{1+\bar{\tau}\gamma(\omega_0)F_{\rm ac}^{\,2}} 
\label{EqInversionSaturation}
\end{equation}
with $\bar{\tau}=\tau_1+  \tau_2 (1-\tau_1/\tau_{21}) \approx\tau_1+  \tau_2 $ as
 the ratio $\tau_1/\tau_{21}$
should be minimized for a good QCL design. Following
Ref.~\cite{WackerBook2012}, we can relate the linear response gain for
a two level system to the $\gamma$-factor
as
\begin{align}
  \label{eq:gainwithgamma}
  G_{12}(\omega) = \frac{2\hbar \omega}{c\epsilon_0\sqrt{\epsilon_r}d} \gamma_{12}(\omega)  \Delta n_{21}\, .
\end{align} 
Following Ref.~\cite{LindskogAPL2014},
Eq.~\eqref{EqInversionSaturation} provides  the
gain saturation
\begin{align} \label{eq:gainfit}
G^\textrm{fit}(F_\mathrm{dc},F_\mathrm{ac},\omega_0,\omega)=  \frac{ G_{12}^{\,0}(\omega)}{1+\bar{\tau}\gamma(\omega_0)F_{\rm ac}^{\,2}} \, .
\end{align}
Here, $G_{12}^{\,0}(\omega)$ is the unsaturated linear response
gain from Eq.~\eqref{eq:gainwithgamma} with $\Delta n_{21}=\Delta n^0_{21}$.
Note, that we have now separated the pumping ac field at
frequency $\omega_0$ from the probe at $\omega$, which means that we
can now probe the saturated gain at a chosen frequency going beyond our NEGF simulations, where only a single frequency $\omega=\omega_0$ appears.

\begin{figure*}
  \includegraphics[width=\textwidth]{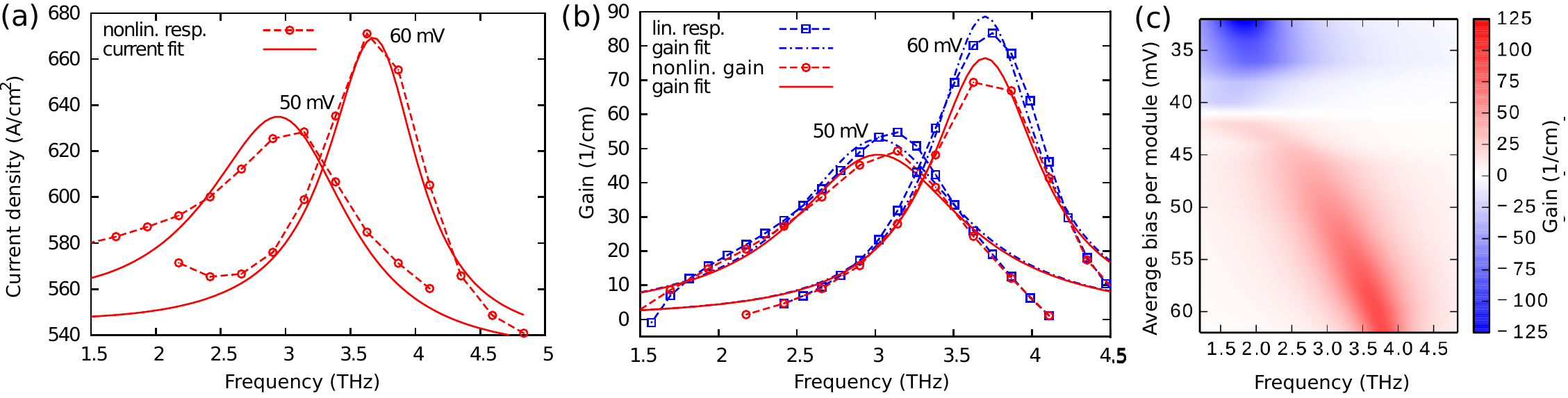}
   \caption{(Color online) In panels {\bf (a)} and {\bf (b)} we compare the NEGF
     results (shown as dashed lines with symbols) with the fitted
     effective expressions for current,
     Eq.~\eqref{eq:jfit} (full lines in {\bf(a)}), 
     and gain, Eq.~\eqref{eq:gainfit} (dash-dotted lines for 
     linear response, full lines for nonlinear response in {\bf(b)}),
     respectively. The non-linear response, i.e. simulations using a
     finite field strength of $eF_{\rm ac}d = 5$ meV, is shown by dashed
     lines with circles whereas linear response calculations, only
     panel (b), are shown with dashed lines with squares. 
     Panel {\bf
       (c)}  shows a map of the gain as a function of frequency and
     bias per module as obtained from $G^\textrm{fit}$ for low
     $F_{ac}$. }
   \label{FigHomsimulation}
\end{figure*}

Under the assumption that the bottleneck for current is the inverted
population at the laser transition, the total current will be a sum of
stationary and induced current following
\begin{align} \label{eq:jfac}
  J(\omega_0) = J_0 + e |n_{2}-n_{1}| \gamma_{12}(\omega_0) F_{\rm ac}^{\,2} 
\end{align}
where only the frequency $\omega_0$, saturating
the system, appears. Here, we use the  absolute value of the inversion,
following our observation, that the
current also increases with absorption in the NEGF simulations, 
due to new transport channels opening up. 
The expression can readily be generalized to many modes of
finite intensity, yielding a sum over field strengths at different
frequencies $\omega_i$, each with $\gamma_{12}(\omega_i)$. 
Identifying
the gain in the expression for current, using
Eq.~\eqref{eq:gainwithgamma}, gives
\begin{align} \nonumber
J^\textrm{fit}(F_\mathrm{dc},F_\mathrm{ac},\omega_0)
 &= J_0 + e d |G_{12}(\omega_0)|\frac{c\epsilon_0\sqrt{\epsilon_r}F_{\rm ac}^{\,2}}{2\hbar \omega_0}\\ \label{eq:jfit}
&= J_0 + e d |G_{12}(\omega_0)|\frac{I(\omega_0)}{\hbar\omega_0} 
\end{align} 
where the total gain $G(\omega)$ contains  a sum over the relevant
transitions. The last factor in the first equation has been  rewritten
as the Poynting vector giving the intensity $I(\omega_0)$ at
$\omega_0$ in the final expression. 

To map the NEGF simulation results onto the effective model,
we first calculate $J_0$ and $G^0(\omega)$ for vanishingly small
$F_{ac}$ at each bias point $F_{dc}d$. We consider the 
range of  35 -- 65 mV per module and frequencies 1.5 -- 4.5 THz
as relevant for domain formation. 
$G^0(\omega)$ is fitted by $\Delta n^0_{21}$ and $\gamma(\omega)$.
According to Eq.~\eqref{eq:gainfit} the
saturated gain can then be described by  the effective lifetime
$\bar\tau$, which is found from fitting the gain spectra to the NEGF
simulations at higher intensities, as exemplified in
Fig.~\ref{FigHomsimulation}(b). We thus require three fit parameters
($\Gamma,z_{21},\bar\tau$), together with the calculated $E_{21}$ and
$\Delta n^0_{21}$ to model the saturated gain and current
at each bias point.  Here, the
NEGF simulation data is fitted to a target function consisting of both
Eq.~\eqref{eq:gainfit} and~\eqref{eq:jfit} which balances both
the gain and  stimulated current contributions and makes efficient use
of all simulation data.  The procedure provides a physically sound way
of interpolating through  the simulation data and allows us to
calculate the current under  irradiation as a function of bias,
frequency and ac field strength.
Representative current and gain spectra are shown in
Figs.~\ref{FigHomsimulation}(a) and (b), respectively, together with
their fits using Eq.~(\ref{eq:jfit}) and Eq.~(\ref{eq:gainfit}).

Figs.~\ref{FigHomsimulation}(a,b) shows a particularly large linewidth
for a bias of 50 mV per module. The reason can be seen in in Fig.~
\ref{FigDensityplots}(b), where the lower laser level and the
extraction level form a doublet with a spacing of 5 meV. Thus the gain
is based on two different transitions of approximately equal strength
resulting in the large linewidth.

Unsaturated gain for a single module as a function of bias and
frequency is displayed 
in Fig.~\ref{FigHomsimulation}(c). A strong Stark shift can be seen
as bias is increasing. At the nominal operation point of 54 mV per
module, gain peaks around 3.3 THz, which agrees with the experimental
lasing frequencies reported in Ref.~\cite{FathololoumiJAP2013}, see
also Fig.~\ref{FigSpectra}. 

%%%%%%%%%%%%%%%%%%%%%%%%%%%%%%%%%%%%%%%%%%%%%%%%%%%%
\section{Experimental methods and measurements} \label{SecSetup}

\begin{figure}
 \centering
 \includegraphics[width=0.8\columnwidth]{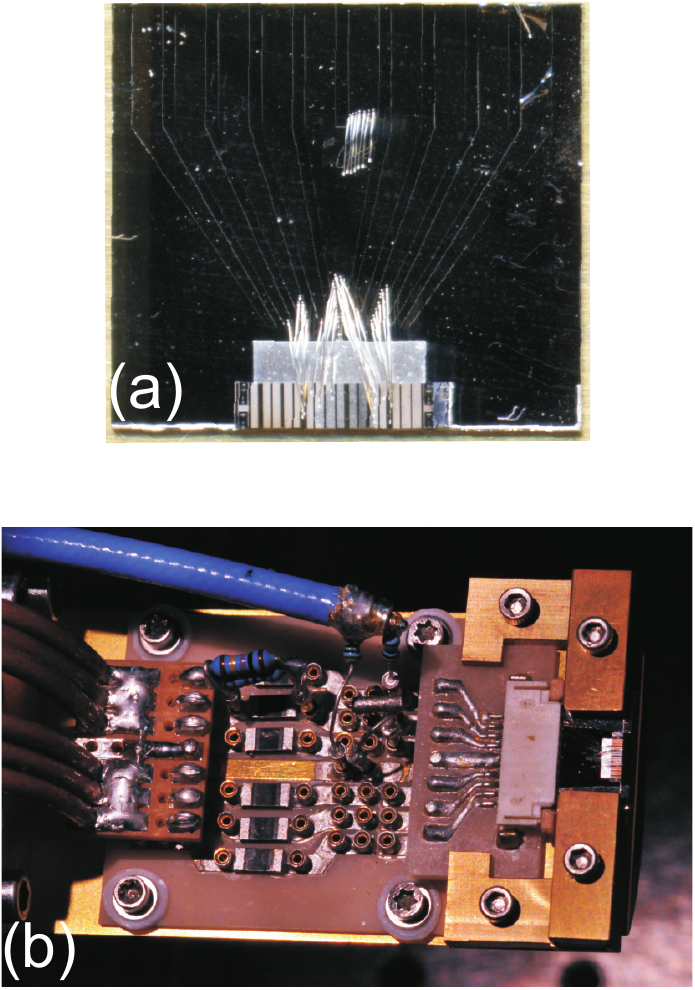}
 \caption{(Color online) Cold finger inside the 10 K closed-cycle refrigerator: {\bf
     (a)} Image of the TiAu metal-metal waveguide laser bar,
   1049~$\mu\rm m$ long, soldered onto a high resistive Si
   carrier. The fan-out electrodes deposited on carrier can be seen,
   but with a low contrast due to their (too) small thickness; this is
   why they showed a residual resistivity. Panel {\bf (b)} The laser
   bar mounted on the cold finger. The blue wire is a 18 GHz SMA cable
   used for voltage detection and on the other side, it is connected
   to a 50 Ohm input channel of a high bandwidth oscilloscope. About
   25 mm are separating the laser bar from the SMA cable with, in
   between, a circuitry that is not optimized for RF.}
 \label{FigColdFingerImage}
 \end{figure}

We consider the structure V812 reported in
Ref.~\cite{FathololoumiJAP2013} using devices processed with a Au-Au
double metal waveguide. These showed lasing up to the heat-sink
temperature of 160 K. The laser under study, 1.049-mm long and
143-$\mu\textrm{m}$ wide, is Indium-soldered onto a highly resistive
($> 10^4$ Ohm.cm) hyper pure floating zone Si carrier on which fan-out
Au electrodes were fabricated, see
Fig.~\ref{FigColdFingerImage}(a). Laser devices and simple mesas (non
lasing) with Pd/Ge/Ti/Pt/Au Ohmic contacts were also fabricated and
mounted on Si carrier. 

It occurred that both lasers, TiAu and PdGe contacted, were mounted on
carriers with too thin electrodes that showed residual, although non
negligible, resistance. The effect of the ground electrode resistance
on measurements could not be circumvented easily, at least not in
pulsed mode operation. Fortunately, the PdGe mesa devices were mounted
on a carrier with much thicker electrodes (hence their vanishing
resistance) and their electrical $J_{\textrm{mesa}}-U_{\textrm{mesa}}$
characteristics served as reference in order to derive the ground
electrode resistance of the Si carrier, the Schottky barrier (for TiAu
only) and therefore the exact bias per module on laser devices
[Fig.~\ref{FigDomainFormation}(a)]. Consequently, in
Fig.~\ref{FigDomainFormation}(a) the experimental average bias per
module for the PdGe contacted mesa device is simply given by the
signal from the voltage probe divided by $N=222$. This is in contrast
with lasing devices, where the potential drop across the ground
electrode resistance of $\sim$1 Ohm (TiAu, dashed green) and $\sim$
0.85 Ohm (PdGe, dashed purple) of their ``defective'' Si carrier, as
well as the Schottky top contact drop of 0.65 V for TiAu device, had
to be taken into account. It is worth mentioning that the PdGe mesa
device was tested in voltage-controlled conditions, i.e. with a small
load resistance $\textrm{R}_L=5$ Ohm, which could explain why the
current plateaus are not very flat (Fig.~\ref{FigDomainFormation}(a),
purple solid line). In Figs.~\ref{FigDomainFormation}(b),~\ref{FigExpOsc}(a,c,e) the ``raw'' QCL voltage is not corrected from the effects of Schottky barrier
and ground electrode resistance of Si carrier; the latter is likely to
be temperature dependent. This explains to a large extend the
difference between the vertical scales of
Fig.~\ref{FigDomainFormation}(b) and (c). Nevertheless, the overall
good agreement between theoretical and experimental biases per module
can be appreciated in Fig.~\ref{FigDomainFormation}(a).

High precision $L-U_{\textrm{QCL}}-J$ characteristics were recorded
using current controlled conditions, by use of a series resistance of
$R_L=41.2$ Ohm placed near the device in a 9 K closed-cycle He
refrigerator system, see Fig.~\ref{FigCircuitNumbering}(a) for the circuit. The external bias $U_0$ is supplied to the
circuit by a pulse generator. The total current in the circuit, $I_{\rm
  tot}$, was measured at the output port of the pulse generator with a
calibrated 200 MHz bandwidth current transformer. Because of the high
resistive Si carrier, the device is operated with a floating
ground. We made sure and checked that the net current in the 50 Ohm
coaxial cable from the voltage pulser is zero, meaning the current
flowing in the QCL returns to the pulser ground via the shield of
coaxial cable.

The voltage across the QCL, $U_{\textrm{QCL}}$, is recorded via a
four-point measurement setup and with a calibrated voltage probe
placed inside the cryostat and terminated to 50 Ohm on the
oscilloscope. The probe consists of a 18 GHz bandwidth SMA cable on
which small footprint 1000 and 330 Ohm resistors were tightly soldered
to center conductor and shield respectively. Despite the high quality
SMA cable used here, the RF performance of the voltage measurement
setup is hindered by the absence of RF design between the input port
of the voltage probe and the device that are $\sim$25 mm apart, see
Fig.~\ref{FigColdFingerImage}(b). Before reaching the voltage probe,
the RF signal has to propagate through simple wirebonds connecting the
QCL electrodes and the tip of fan-out electrodes patterned on Si, then
through non-RF designed fan-out electrodes on the Si carrier, then
conductors on two AlN boards and finally the resistors used for the
voltage probe. This incomplete RF design results in a ringing effect
that is partially filtered out in Appendix~\ref{SecDeconv}. In the
future, a proper RF designed setup to probe the QCL voltage should be
considered.

The small bypass current in the voltage probe (see Fig.~\ref{FigCircuitNumbering}(a)), $I_{R_p}$ was
subtracted in the final reported results of current density in
Fig.~\ref{FigDomainFormation}(b), $ J\equiv (I_{\rm tot}-I_{R_p})/A$
in DC mode, and its resistance, $R_p$, was taken into account in the
load curve that is explained with Eq.~\eqref{loadline} in next
section, see also Fig.~\ref{FigCircuitNumbering}(a). The
total circuit current $I_{\rm tot}$ and the QCL voltage $U_{\rm QCL}$
were  measured in pulsed mode with 2-$\mu$s pulses, repeated at 100
Hz frequency and with 14 Hz macro modulation of the train of pulses to
adapt the measurement to the response time of the Golay cell, the THz
detector. The output of the Golay cell was connected to a lock-in
amplifier with 1 s time constant and locked to 14 Hz reference
signal. Here, unusual long pulses were used to benefit from the good
flatness of these pulses in their second half and consequently report
high accuracy data and longer sections of oscillations, when they
occur. Small heating effects on laser threshold have been identified
and found responsible for vertical ($U_{\rm QCL}$) shift of the merlon
for long pulses, the shift depending on the position of integration
window where current and voltage are monitored. Experimentally, we
observed the merlon at lower bias, if shorter pulses are applied
(about 0.1 V lower for 300-ns pulse duration).
This is why the electrical characteristics (Fig. 2(a, b) for instance) were recorded near the center of the 2-$\mu$s pulse, not close to the trailing
edge.
A 5 GS/s, 1 GHz bandwidth digital oscilloscope (Tektronix DPO4104) was
used to zoom and measure the top of the pulses $I_{\rm tot}$ and
$U_{\rm QCL}$ by using appropriate DC offsets and vertical scales. The
$I_{\rm tot}$ and $U_{\rm QCL}$ values were measured inside a 200-ns
long integration window defined between cursors placed at
$t=1.08~\mu\rm s$ and $t=1.28~\mu\rm s$ inside the pulse. Sixty four
current and voltage traces were averaged (average mode of
oscilloscope) and finally reported with 1 and 2 mV/div scales
respectively, meaning that any changes of vertical scale that can
result into discontinuities were taken into account and thoroughly
corrected. Before turning on the average mode on the oscilloscope,
clipping the single pulse signals (observed in sample mode)  by the 10-division vertical dynamic range
was carefully avoided in the integration window (only). The calibration of the pulse generator was also checked
as well as that of the oscilloscope. Data from oscilloscope and lock-in
amplifier were recorded 32 times, 1 s apart, and finally means and
standard deviations were derived. We used 50 mV voltage steps on the
pulse generator in order to resolve features on the merlon. At each
step, we also recorded 200 single traces, i.e. non averaged (sample
mode), of $I_{\rm tot}$ and $U_{\rm QCL}$ channels to capture any
oscillations or discontinuities. The median value of peak-to-peak
amplitude of these pulses was computed inside the same 200-ns window
of interest. Moreover, this median value was subtracted from the
peak-to-peak amplitude of 64-pulse averaged signal to account for non
perfect flatness of voltage and current channels in the integration
window. After this subtraction the final value was named the ``net''
amplitude, which is shown for the entire temperature range in
Fig.~\ref{FigExpOsc}(a) superimposed on the bias signal as shaded
regions. Note, that these amplitudes are only qualitative due to the
  limited bandwidth of the oscilloscope used here.
When voltage oscillations were detected, current oscillations
were also observed albeit less intense due to limited bandwidth of
current transformer.

The time-resolved voltage oscillations were recorded with a 40 GS/s 15
GHz bandwidth real-time oscilloscope (Tektronix TDS6154C) and a 10 dB
attenuator was employed due to the large oscillation amplitude in
region III. The first set of measurements with the 1 GHz oscilloscope
permitted to locate the oscillations with a low vertical resolution as
a function of external bias. With the help these low resolution
``spectra'' we drove the time-resolved measurements with TDS6154C for
different external biases and temperatures; single pulses were sampled
at a maximum resolution, 25 ps/point.

Due to the residual positive slope of the external bias from the
pulser ($U_0$), single pulse measurements could reveal several
sections of the merlon at once (such as regions III and IV). This
simple technique, described in Appendix~\ref{AppBiasDependence}, was
convenient to assess the oscillating frequency versus raw QCL voltage.

The spectral measurements were performed with a continuous scanning
Fourier transform infrared spectrometer set at a mirror velocity speed
of 0.05 cm/s and resolution of 0.1 $\textrm{cm}^{-1}$ and equipped
with a LHe cooled Si bolometer. We used a double modulation technique
in which the QCL was biased with 0.5-$\mu\rm s$ pulses at a repetition
rate of 300 Hz. The signal from bolometer was filtered with high pass
200 Hz filter and finally sent to lock-in amplifier with time constant
of 10 ms. Only five or ten scans were accumulated by FTIR.

In passing we also note that high resolution X-ray diffraction
measurements indicate that the experimental layers are 1.7\%
shorter. For thinner wells, the tunnel resonances are shifted to
larger biases and the lasing transitions to higher frequencies, which
might reflect some minor discrepancies observed between theory and
experiment.

%%%%%%%%%%%%%%%%%%%%%%%%%%%%%%%%%%%%%%%%%%%%%%%
\section{Additional modeling results}
\label{AppAddResults}

\begin{figure*}
  \centering
  \includegraphics[width=\textwidth]{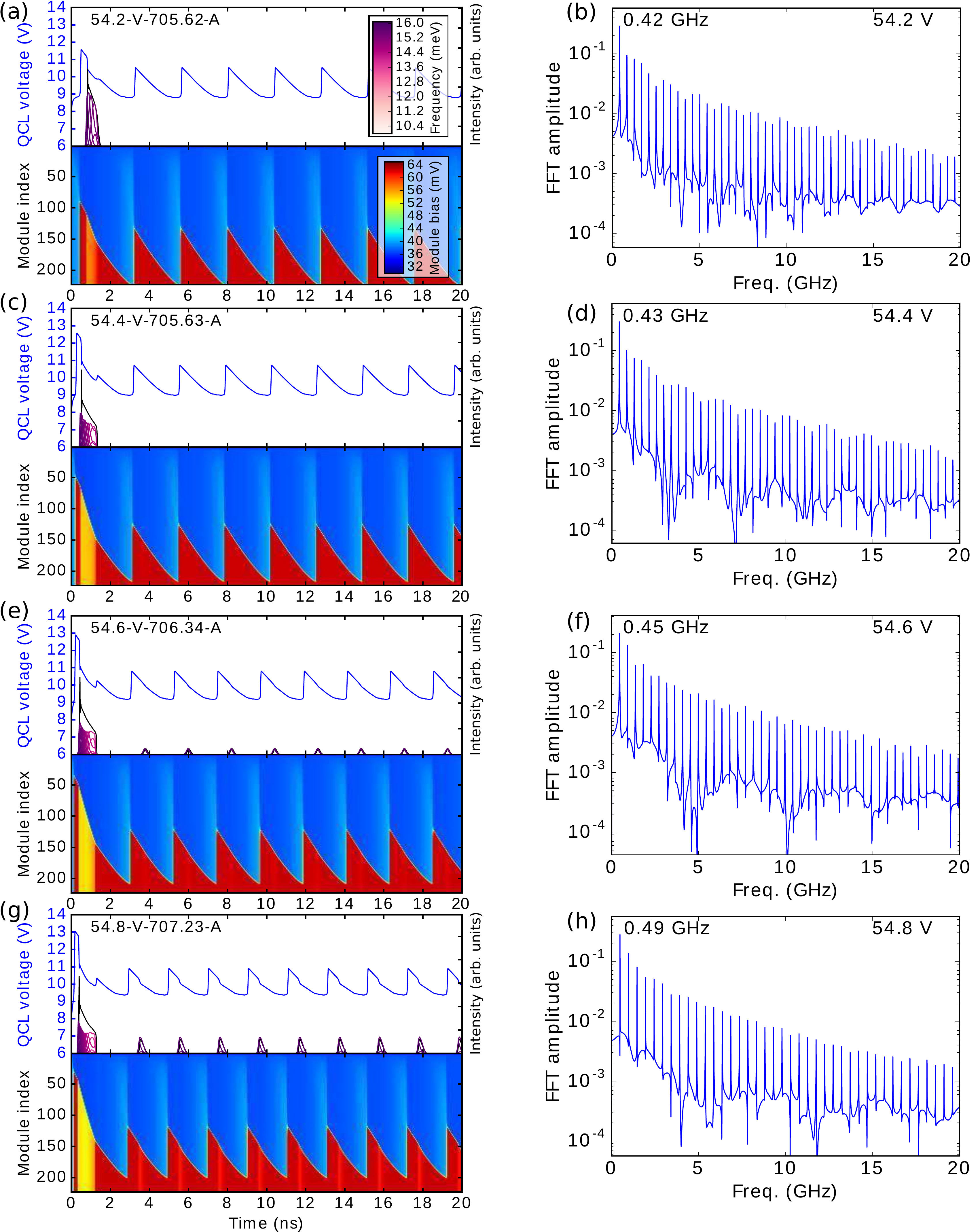}
  \caption{(Color online) Traveling domains simulations for external biases
    54.2-54.8 V as indicated in the panels, to complement the data in
    Fig.~\ref{FigSimOscRegions}. The color bars in the top and bottom
    panels of the left section apply also to the rest of the column,
    where they have been removed for increased clarity. Fast Fourier
    transforms (FT) of the simulated voltage oscillations $U_{\rm
      QCL}(t)$ are shown at matching external bias in the right
    column. The voltage oscillations for the FT were recorded over a
    sample time of 200 ns. In each such panel, the frequency of the
    strongest harmonic is indicated. }
  \label{fig:sim_osc_fft_0}
\end{figure*}

\begin{figure*}
  \centering
  \includegraphics[width=\textwidth]{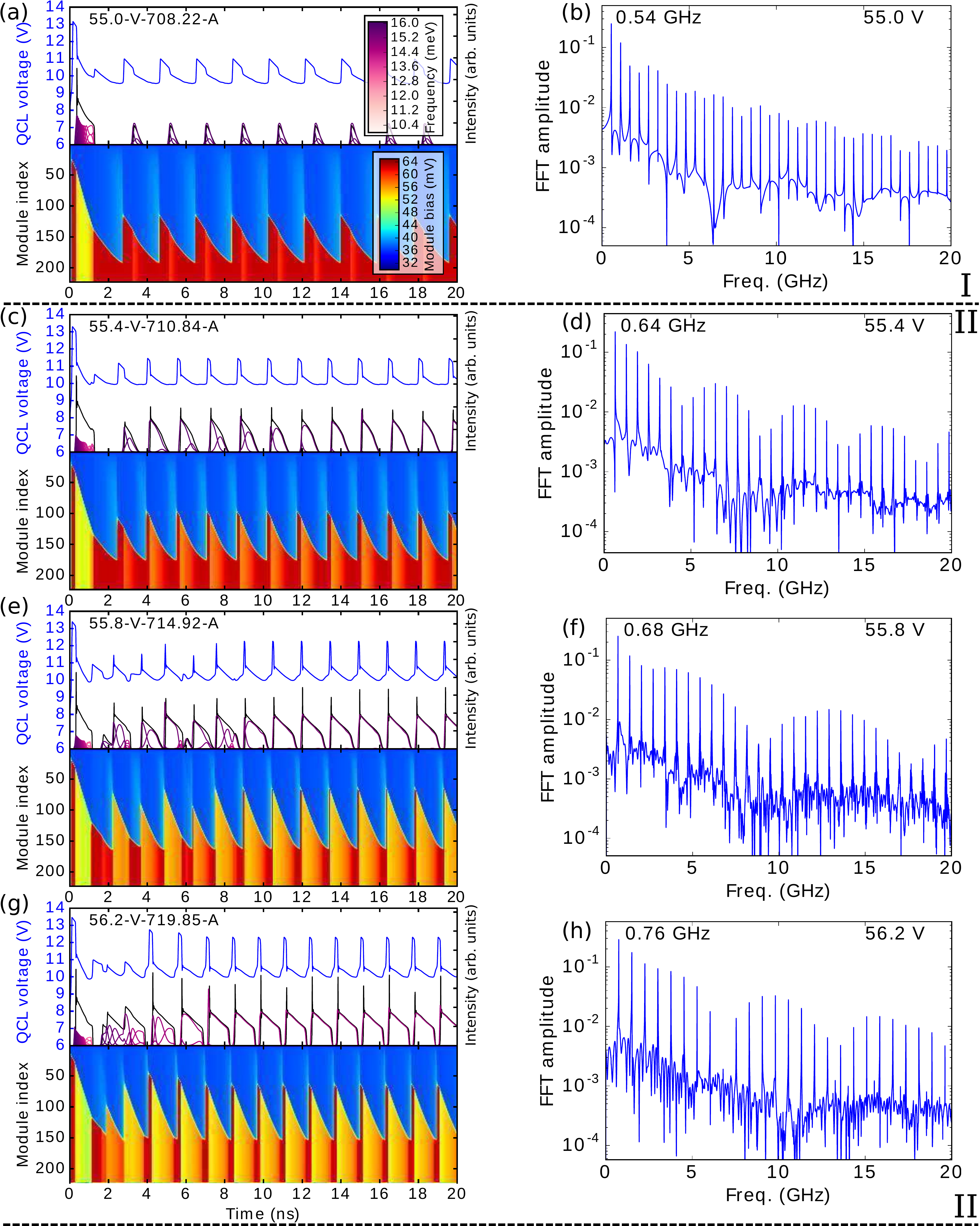} 
  \caption{(Color online) Traveling domains simulations for external biases
    55.0-56.2 V as indicated in the panels, to complement the data in
    Fig.~\ref{FigSimOscRegions}. The color bars in the top and bottom
    panels of the left section apply also to the rest of the column,
    where they have been removed for increased clarity. Fast Fourier
    transforms (FT) of the simulated voltage oscillations $U_{\rm
      QCL}(t)$ are shown at matching external bias in the right
    column. The voltage oscillations for the FT were recorded over a
    sample time of 200 ns. In each such panel, the frequency of the
    strongest harmonic is indicated.}
  \label{fig:sim_osc_fft_1}
\end{figure*}

\begin{figure*}
  \centering
  \includegraphics[width=\textwidth]{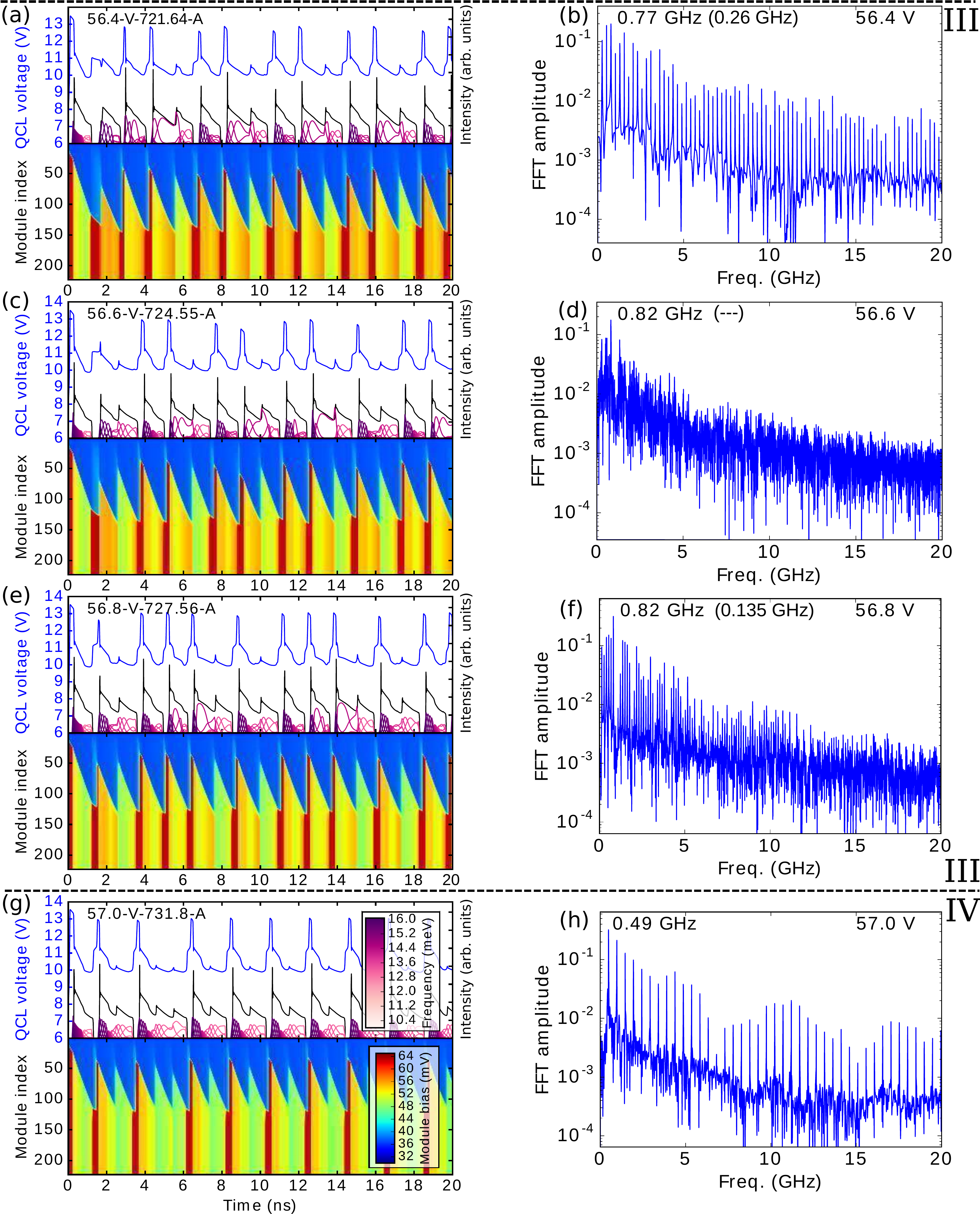} 
  \caption{(Color online) Traveling domains simulations for external biases
    56.4-57.0 V as indicated in the panels, to complement the data in
    Fig.~\ref{FigSimOscRegions}. The color bars in the top and bottom
    panels of the left section apply also to the rest of the column,
    where they have been removed for increased clarity. Fast Fourier
    transforms (FT) of the simulated voltage oscillations $U_{\rm
      QCL}(t)$ are shown at matching external bias in the right
    column. The voltage oscillations for the FT were recorded over 
    sample time of 200 ns. In each such panel, the frequency of the
    strongest harmonic is indicated. If the fundamental
      frequency differs, its value is added inside brackets.}
  \label{fig:sim_osc_fft_2}
\end{figure*}

\begin{figure*}
  \centering
  \includegraphics[width=\textwidth]{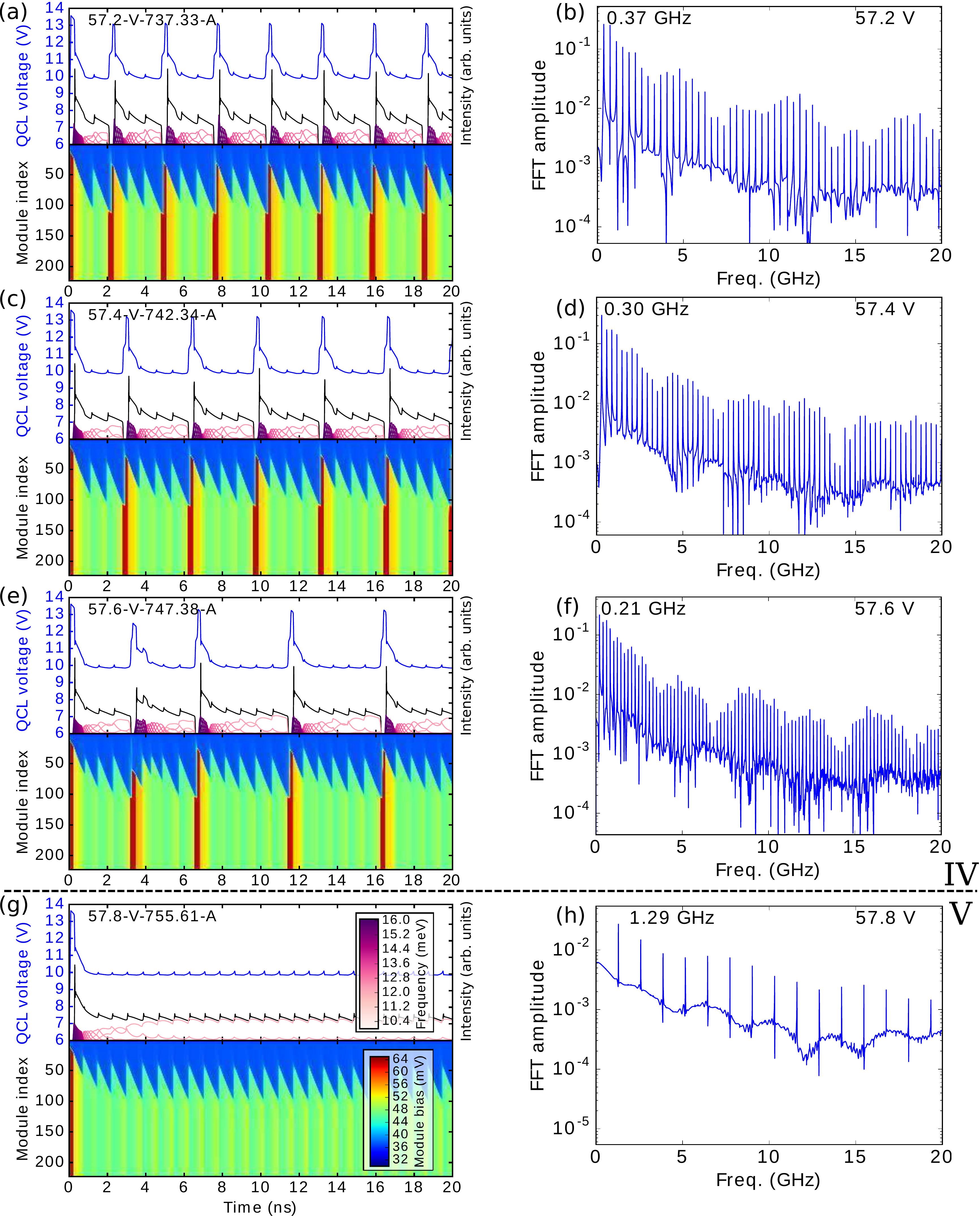} 
  \caption{(Color online) Traveling domains simulations for external biases
    57.2-57.8 V as indicated in the panels, to complement the data in
    Fig.~\ref{FigSimOscRegions}. The color bars in the top and bottom
    panels of the left section apply also to the rest of the column,
    where they have been removed for increased clarity. Fast Fourier
    transforms (FT) of the simulated voltage oscillations $U_{\rm
      QCL}(t)$ are shown at matching external bias in the right
    column. The voltage oscillations for the FT were recorded over a
    sample time of 200 ns. In each such panel, the frequency of the
    strongest harmonic is indicated. }   
  \label{fig:sim_osc_fft_3}
\end{figure*}

To complement the data shown in Fig.~\ref{FigSimOscRegions}, a set of
plots detailing the domain formation simulation results are shown in
Figs.~\ref{fig:sim_osc_fft_0}-\ref{fig:sim_osc_fft_3}. They describe
the simulated oscillations from just before threshold to a highest
current density of 1.07 $J_{\rm thr}$, corresponding to a change in
external bias, $U_0$, from 54.2 V to 57.8 V, respectively. At the high
end of the current, see Fig.~\ref{fig:sim_osc_fft_3}(g, h), the lasing
is stabilized in the sense that the high-field domain always stays
close to 50 mV/per period and the laser action is
uninterrupted. Increasing the external bias further, this state  will
eventually merge with the homogeneous branch as shown in 
Fig.~\ref{FigDomainFormation}(a). 

Figure \ref{fig:sim_osc_fft_0} focuses on the oscillating domains of
region I. Here the oscillation frequency is slowly increasing with
bias as the new instability occurs at a larger extend of the previous
high-field domain. Thus, the accumulation front travels a shorter
range, resulting in the frequency increase with $U_0$. The Fourier
transforms (FTs) in the right hand column were taken over a long
signal (200 ns) compared to short stretch of time of only 20 ns used
for the voltage and module color plots on the left side. In each of
the panels the frequency of the strongest oscillating signal is given
together with the lowest frequency distinguishable in brackets. In
panels (e, g) threshold is actually reached for short periods in time,
but the lasing field is too weak to influence the local current-field
relation. Thus the high-field domains keep the field strength of
$\sim$62 mV per module.

Figure \ref{fig:sim_osc_fft_1} shows the transition from region I to
region II, where the lasing strongly increases and affects the
high-field domains, whose fields drop. Due to the Stark shift, this
implies a drop in lasing frequency, and lasing is now also observed
for lower frequencies, see also Fig.~\ref{FigSpectra}. However,
lasing stops, after the high-field domain has shrunk sufficiently due
to the traveling accumulation front.  On the right hand side, we note
an increase in the oscillation frequencies. 

The next collection of plots in Fig.~\ref{fig:sim_osc_fft_2} shows the
region III and the transition to region IV. Here the lasing is still
active while some new high-field domains form with a field of $\sim
$50 mV per module, i.e. close to the NOP. Thus two different domain
formation scenarios coexist. These do either arrange in a pattern such
as in panels (a, b) and (e, f) with subharmonic frequencies $f/3$ and
$f/6$, respectively, or become irregular as seen in panel (c, d). A
similar subharmonic feature could also be detected experimentally, see
Figs.~\ref{FigExpSubharmonics} and \ref{FigDeconvExample}(b) below.
Within region III, where the average
bias drop over the QCL is roughly constant, less than half of the new
high-field domains exhibit this low field. For the lowest panels (g,
h), the two different domain formation processes alternate. For this
operation point, the average bias reaches a maximum, defining the
transition to region IV.

In the final collection of plots in Fig.~\ref{fig:sim_osc_fft_3}
decreasing oscillation frequencies of region IV can be studied, where
the interruptions in lasing become less and less frequent. As
field-domains with high field become rare, the average bias drops in
this region upon increase of the total bias $U_0$, which defines
region IV. In the lowest panels (g, h) the lasing is uninterrupted
and region V is reached, where the laser operates with a linear
increase of bias and intensity with current. Our simulations show some
oscillatory feature. However the amplitude of the bias oscillation is
strongly reduced compared to the other regions.

\section{Bias dependence of oscillation frequency}
\label{AppBiasDependence}

The voltage dependence of the oscillation frequency can be
demonstrated by repeating the measurement of the raw QCL voltage with
the 15 GHz bandwidth oscilloscope at different external
biases $U_0$. However, this obvious measuring technique was hampered by the
pulse-to-pulse variations. Therefore, a faster method was also
employed which relied on the residual positive slope of the voltage
input pulse and a local Fourier transform within a
Gaussian window covering $\sim$40 ns. This slope is non negligible 
for $t<1~\mu\rm s$ and is greatly reduced at longer time as the 
external bias eventually converges to the setpoint value.

In the set of experiments reported in the
Appendices~\ref{AppBiasDependence} and \ref{SecDeconv} we used
actually a different device than the one in Sec.~\ref{SecDomain}
and \ref{SecOscillations}.  This device is from the same laser bar,
mounted on the same Si carrier, and its electrical characteristics,
the merlon for instance, are very similar. However, the bias
oscillations were stronger with a peak-to-peak value up to  3.6 V as
monitored by the 15 GHz bandwidth oscilloscope.

\begin{figure*}
  \centering
  \includegraphics[width=0.77\textwidth]{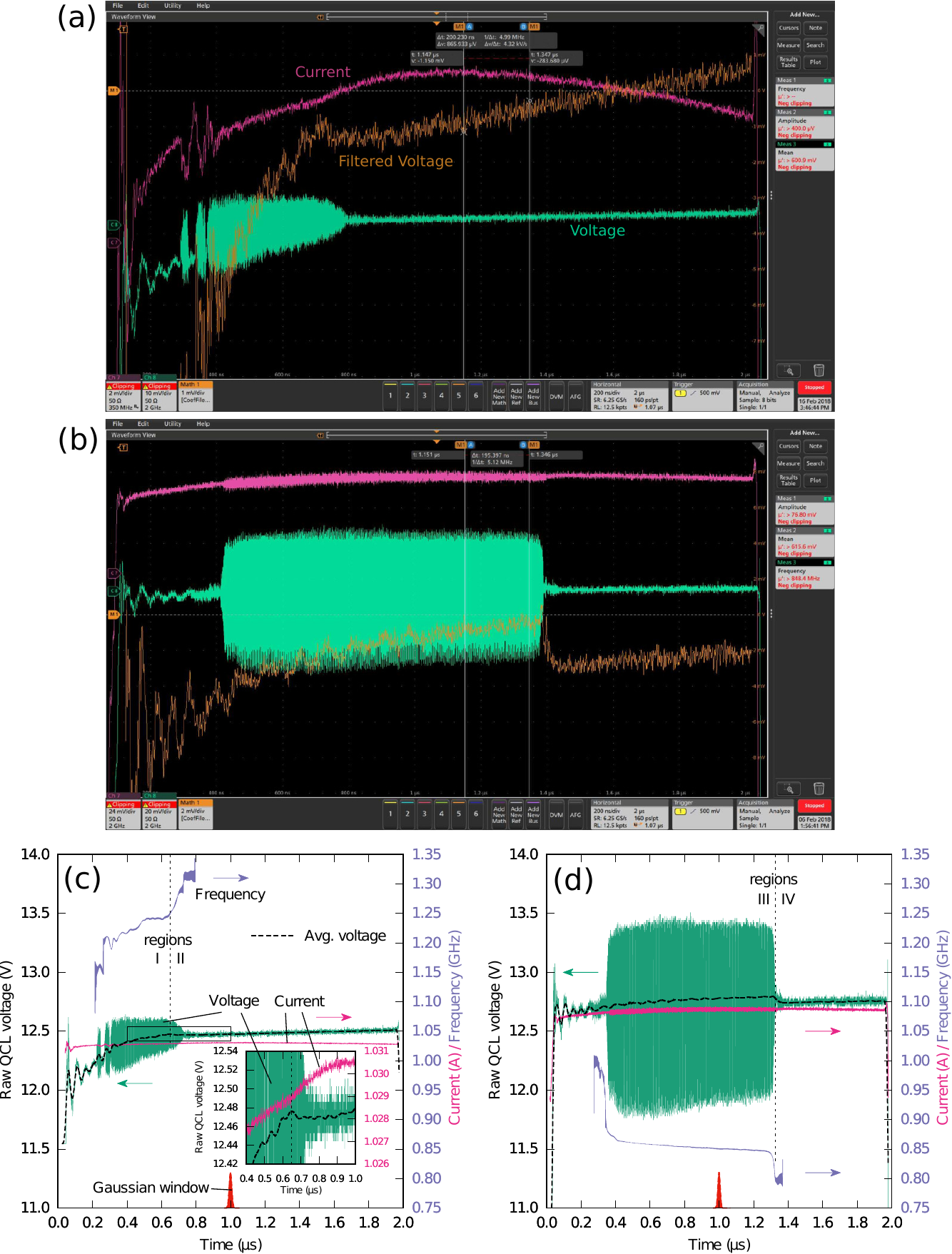}
  \caption {(Color online) Time-resolved oscillations at 9 K over the 2-$\mu $s long
    pulse with a residual positive slope as measured with a 2 GHz
    bandwidth oscilloscope. Panels {\bf (a, b)} are screen captures of the
    oscilloscope for $U_0=55.24$ V (i.e. around lasing threshold) and
    $U_0=57.9$ V respectively. As also indicated in {\bf(a)}, the magenta trace represents the
    current ($I_\mathrm {tot}$) pulse, the green trace the raw QCL
    voltage and the orange trace shows the QCL voltage channel with a
    312.5 MHz low-pass filter. The sharp step down on the filtered
    voltage in panel (b) illustrates the right side of the merlon,
    region IV, previously recorded in average mode 
    [see Fig.~\ref{FigDomainFormation}(b)]. Panels {\bf (c, d)} provide a detailed analysis of the data
    from (a, b), respectively. The purple curve (labeled \textit{Frequency}) represents the
    instantaneous 1st harmonic frequency computed within a moving
    integration window, which is represented by a filled area red
    curve (labeled \textit{Gaussian window}) at the bottom. 
    The dashed black line represents the voltage
    averaged within the Gaussian window. The inset in panel (c)
    focuses on the identification of threshold, the boundary between
    regions I and II, at $t\approx 0.65~\mu $s: (i) The average slope
    of the QCL voltage decreases; (iii) The average slope of the QCL
    current increases slightly; (iii) The voltage exhibits a tiny
    spike of 10 mV at  $t\approx 0.65~\mu\mathrm{s}$, similar to Fig.~2(b)
    at threshold (where 20 mV is observed in average mode). The vertical dashed lines represent region
    boundaries, which we could clearly identify. The slow
      decline of current and increase of average voltage for $t\gtrsim
      1.3\, \mu\mathrm{s}$ are attributed to sample heating.}
    \label{FigFreqVsTime1}
  \end{figure*}

For the data reported here, we employed a 6.25 GS/s, 2 GHz bandwidth
oscilloscope (Tektronix MSO58) with its 8-bit analog-to-digital
converter to benefit from its maximum sampling rate. A digital
low-pass filter with a cutoff frequency of 312.5 MHz was available,
which allowed the real-time visualization of changes in average bias as
characteristic for different regions in the merlon. By carefully
adjusting the setpoint on the pulser, we could observe the
oscillations in region I with the transition to region II, see
Fig.~\ref{FigFreqVsTime1}(a), as well as the oscillations in region
III with the transition to region IV, see Fig.~\ref{FigFreqVsTime1}(b).
Fine tuning of the pulser setpoint and a ``judicious'' choice of the
pulses --due to inevitable pulse-to-pulse fluctuations-- were
necessary to obtain this data.

Figure~\ref{FigFreqVsTime1} shows two examples of oscillations at 9 K
for external bias corresponding to regions I [nominal pulse amplitude
  $U_0=55.24$ V, panels (a, c)] and III [$U_0=57.90$ V, panels (b,
  d)]. The panels (a) and (b) are the oscilloscope screen captures of
the total current pulse (magenta), the raw QCL voltage (green)  and
the low-pass filtered voltage (orange).  The bottom panels (c) and (d)
show the results of a moving Fourier analysis performed on the raw QCL
voltage. For that purpose, a 45 ns (panel (c)) and 40-ns (panel (d)) wide Gaussian windows with a center shifted from the leading to the
trailing edge of the pulse were used. The standard deviation of the Gaussians were 7.5 ns and 6.7 ns respectively. Doing so, in a single shot, the average
voltage within the Gaussian window (black line) can be computed and
the voltage sensitivity of the first harmonic frequency (purple line)
estimated, albeit the non flatness of external bias pulse might
slightly influence the oscillation frequency. 

In Fig.~\ref{FigFreqVsTime1}(a, c) we identify the transition between
regions I and II at $t\approx 0.65~\mu\textrm{s}$ by the sudden change
in slope in bias, which stops its strong increase. More detailed, the
inset of panel (c) displays a small spike in bias ($\sim$10 mV),
similar to the one observed at threshold for low temperatures in
Fig.~\ref{FigDomainFormation}(b) where the spike is $\sim$20 mV in average
mode. Furthermore, a slight increase in the current slope also
indicates the transition to a region with higher average conductivity
as common for region II due to lasing. These features are similar for
all pulses analyzed. Regarding the lasing signal, we observed that
the phase of the lock-in amplifier starts to be stabilized, when the
oscillations vanish at the end of the pulse (strictly speaking for the
majority of pulses with nominally equal set point bias
$U_0$). Fig.~\ref{FigFreqVsTime1}(c), which is representative for many
pulses analyzed, shows that the oscillations are decaying in time
after entering the region II. This relates the transition between
region I and II with threshold, see also Fig.~\ref{FigDomainFormation}(b). 
Therefore, laser threshold occurs in a state of running
electric field domains, a likely hypothesis that could be confirmed
with a fast THz detector~\cite{ScheuringTransTerahertzSciTechnol2013}.
Consequently, for $t< 0.65~\mu\textrm{s}$ the device is operated in
region I, where we observe oscillations with a peak-to-peak amplitude
of $\sim$0.4 V and frequencies of $\sim$1.2 GHz, slightly increasing
with bias. These values agree well with the data in Fig.~\ref{FigExpOsc}(b, c), 
where a different device was studied with a 15 GHz
oscilloscope.

\begin{figure*}
  \centering
  \includegraphics[width=0.78\textwidth]{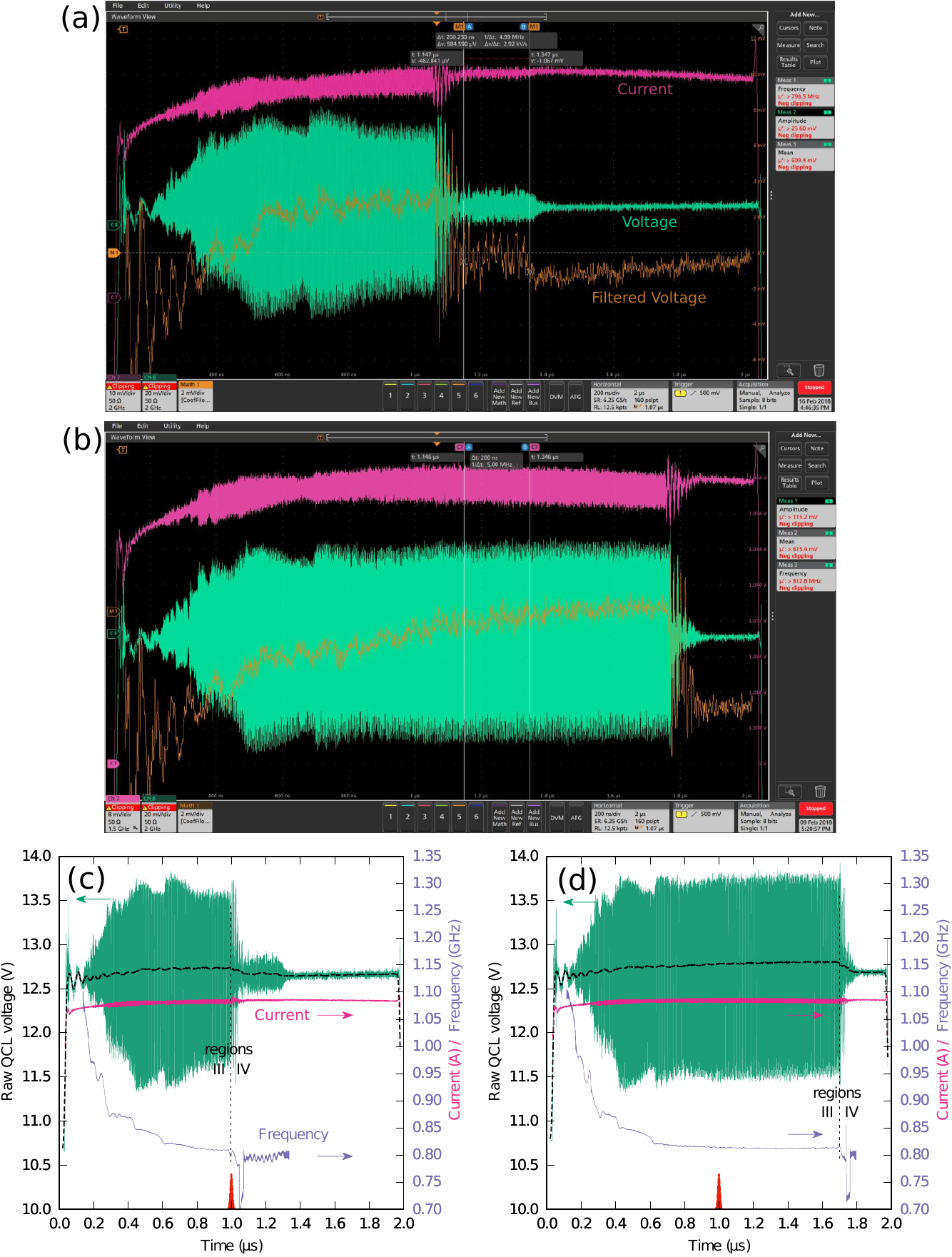}
  \caption{(Color online) Time-resolved oscillations recorded with the 2 GHz
    oscilloscope at 9 K and $U_0=57.76$
    V for two different pulses. All curves correspond to Fig.~\ref
    {FigFreqVsTime1}. The deviations between panels (a, c) and (b, d)
    show typical pulse-to-pulse fluctuations. In both cases a
    multi-segment region III and an instantaneous oscillation
    frequency evolving like a staircase are observed. The $\sim$0.1 V
    drop of average voltage (dashed black line) marks the region IV, the
    right side of merlon in Fig.~\ref{FigDomainFormation}(b).}
    \label{FigFreqVsTime2}
  \end{figure*}
  
\begin{figure*}
  \centering \includegraphics[width=0.8\textwidth]{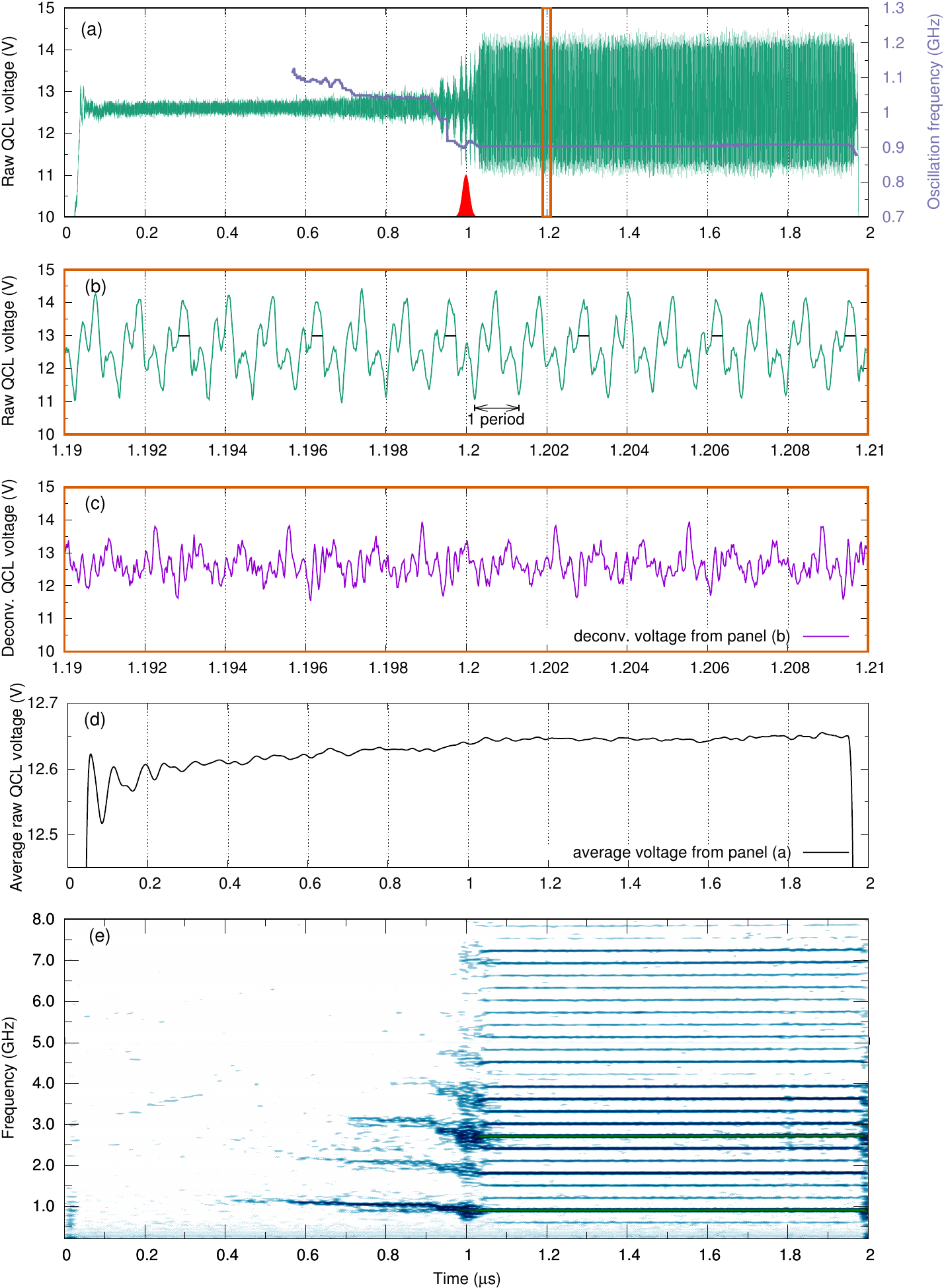}
  \caption{(Color online) Time-resolved oscillations recorded with the 15 GHz
    oscilloscope at 9 K and $U_0=57.54$ V for a pulse with good
    stability. Panel {\bf (a)}, left axis, shows the top of the
    2-$\mu\mathrm{s}$ long pulse of the raw QCL voltage. The right
    axis is the instantaneous frequency as computed within a 50-ns
    wide Gaussian window with 8.5-ns standard deviation displayed
    around t=1 $\mu\mathrm{s}$. Panel {\bf (b)} shows details of the
    recorded bias signal in a 20-ns wide window. The dominant
    frequency of 0.9 GHz corresponds to the period indicated. However,
    by looking into more detail, only one of three periods are
    identical, as it can be seen by the wider peaks marked with a
    black bar. Panel {\bf (c)} shows the same 20-ns wide window after
    filtering out the ringing effects.  Panel {\bf (d)} shows that the
    $f/3$ subharmonic regime is observed when the average raw voltage
    is stabilized at $\approx$12.65 V. Panel {\bf (e)} shows the
    instantaneous frequency spectrum.}
  \label{FigExpSubharmonics}
\end{figure*}

In Fig.~\ref{FigFreqVsTime1}(b) the sharp drop of filtered voltage
(orange) indicates region IV, the only region where the bias over the
QCL drops. This allows the identification of region III for times
$t<1.3\, \mu$s, see also panel (d). Within this region large amplitude
voltage oscillations with peak-to peak amplitude of $\sim$1.8 V and a
frequency of  $\sim$0.85 GHz, decreasing with bias, are detected by
the 2 GHz oscilloscope (with the 15 GHz oscilloscope higher amplitudes
are resolved). These values are again comparable to Fig.~\ref{FigExpOsc}(d, e). 
The oscillations are also  echoed in the current signal
despite the current transformer being limited to a 200 MHz bandwidth.
We observe, that this region is very sensitive to the bias pulses as
shown in Fig.~\ref{FigFreqVsTime2}. The envelope of oscillations
varies greatly from pulse-to-pulse and we see variations from
day-to-day. Nevertheless the key issues are pertinent: i) the giant
amplitude of oscillations of several volts, ii) the oscillation
frequency consistently in $\sim$850 MHz range\footnote{The 
same $\sim$850 MHz oscillation frequency was automatically computed 
by the oscilloscope between the vertical cursors and displayed as 
``Meas 3'' on the right side of the screen capture in Fig. \ref{FigFreqVsTime1}(b).} and iii) an
averaged voltage drop $\sim 0.1$ V within the region IV. While the
voltage drops, the oscillations actually persist, albeit with reduced
amplitude. This is best seen in Fig.~\ref{FigFreqVsTime2}(a, c).

It is noteworthy that the pulse-to-pulse variations sometimes lead to
a region III consisting of several segments with different amplitudes
and with decreasing center frequencies, see
Fig.~\ref{FigFreqVsTime2}(c, d).  The moving FT results in an instantaneous
frequency versus time that looks like a staircase, where the last step is
centered at $\sim$813 MHz. When such a multi-segment region III appears it is
also concomitant with a deeper average voltage step in region IV.

Analyzing the operation point with a raw QCL voltage of 12.65 V in
region III in more detail with the 15 GHz oscilloscope, we could
actually observe subharmomics $f/3$, see
Fig.~\ref{FigExpSubharmonics}. For larger values of $U_0$, we observe
this $f/3$ subharmonic regime systematically, when the QCL bias crosses
the 12.65 V level.

%%%%%%%%%%%%%%%%%%%%%%%%%%%%%%%%%%%%%%%%%%%%%%%%%%%%
\section{Filtering the oscillation patterns by deconvolution} \label{SecDeconv}
For a better interpretation of the large amplitude oscillations of the
raw QCL voltage observed in region III (see Fig.~\ref{FigExpOsc}(e)
and Figs.~\ref{FigFreqVsTime1}-\ref{FigExpSubharmonics}), the step
response of the voltage probe setup was measured by employing a 200-ps
rise-time pulser. In that way the setup was characterized up to
$\sim$7 GHz. During this test, the load resistance $R_L$ was removed
and the bias setpoint $U_0\approx 6$ V was chosen to match the 50 Ohm
DC impedance on the QCL at this bias. This measurement revealed a
general ringing enveloppe at  $\sim$0.93 GHz with a exponential decay
of 1.1 ns, but also some complex features inside this enveloppe. From
classic Fourier analysis we could reconstruct an \emph{estimated}
impulse response of the voltage detection setup. The impulse response
shows complex ringing effects that we could not associate with simple
electrical circuits. This response lasts for $\approx$3 ns and it is
peaked at $t=0.75$ ns after the impulse. When integrating the impulse
response, which is equivalent to derive the step response, the ringing
effect would cause a strong 75\% voltage overshoot at $t=0.87$
ns. With such an impulse response and by employing standard Fourier
analysis we tried to perform deconvolution on the complex oscillation
traces of the raw QCL voltage. Nevertheless, the results of this
operation should be treated with cautiousness considering i), the RF
circuit of the setup is far from being optimized as already explained
in Appendix~\ref{SecSetup} and illustrated via
Fig.~\ref{FigColdFingerImage} and ii), our characterization technique
is limited to $\sim$7 GHz so far.  

\begin{figure}
\includegraphics[width=1\columnwidth]{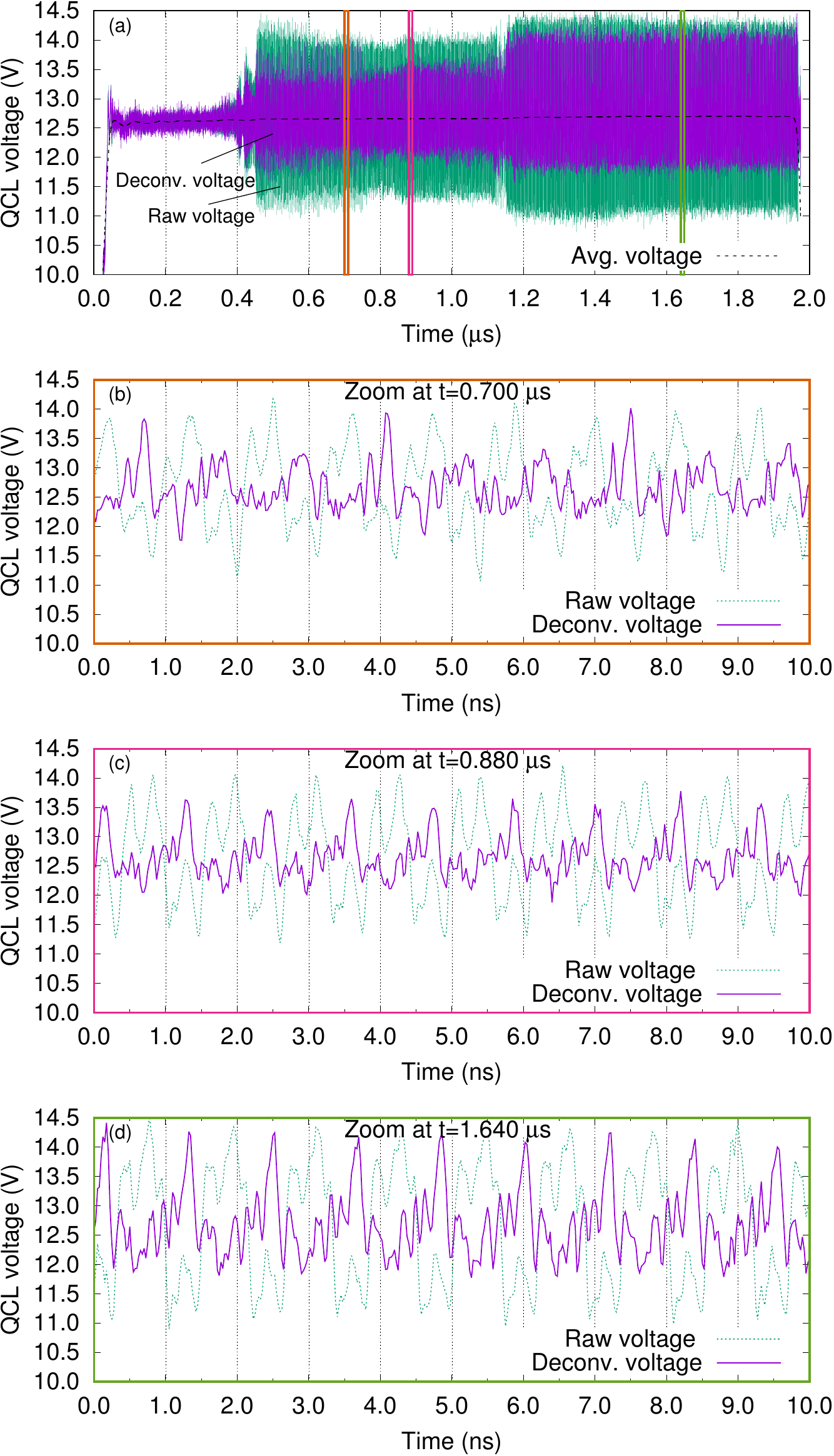}
\caption{(Color online) Exemplary deconvolution of the time-resolved oscillations at
  9 K and $U_0=57.65$ V: {\bf (a)} The 2-$\mu \mathrm{s}$ long green
  trace is the raw signal from the voltage probe recorded with a 15
  GHz bandwidth oscilloscope and the magenta trace with lower
  amplitude oscillations is the result from deconvolution. The dashed black
  trace is the average voltage computed within a moving 50-ns wide
  Gaussian average window (with 8.5 ns standard deviation). A
  multi-segment region III of oscillations can be observed. Panels
  {\bf (b)}-{\bf (d)} represent three 10-ns wide zoomed areas. Panel
  {\bf (b)} corresponds to the $f/3$ subharmonic regime, which becomes
  more distinct after deconvolution.}
\label{FigDeconvExample}
\end{figure}

Figure~\ref{FigDeconvExample} shows the result of deconvolution for an
external bias corresponding to region III. The span of oscillations
does not last 2 $\mu \mathrm{s}$ because of the residual positive
slope of the input pulse (see Appendix~\ref{AppBiasDependence}).
Panel (a)
demonstrates a reduction of the peak-to-peak amplitude oscillation
after deconvolution, suggesting the ringing effects have been filtered
out to a large extend. The next three panels show 10-ns wide zoomed
areas representing  segments, with different
oscillation amplitudes in panel (a).  From a bird's eye view the raw
QCL voltage patterns in the different segments look similar
as ringing effects dominate; nevertheless, after deconvolution the
differences become more evident. When the average voltage is
$\sim$12.65 V we systematically observed a $f/3$ subharmonic regime,
where $f=0.884$ GHz is the main oscillation frequency
(Fig. ~\ref{FigDeconvExample}(b) and Fig.~\ref{FigExpSubharmonics}). The
filtered trace seems to indicate there could be one strong ``spike''
of voltage once every three periods. Such complex oscillations with
three different spikes in a $3/f$ lapse of time, have been predicted
by our model, see Fig. ~\ref{fig:sim_osc_fft_2}(a-b), and our filtered
voltage trace resonates rather well with these simulations. Panels (c)
and (d) show the filtered QCL voltage around $t=0.88$ and $t=1.64$
$\mu \mathrm{s}$ respectively. The average voltage and fundamental
frequency are 12.66 V and 0.874 GHz respectively around $t=0.88$ $\mu
\mathrm{s}$ and 12.7 V, 0.850 GHz around $t=1.64$ $\mu
\mathrm{s}$. These filtered traces display general square-like $\sim
0.5$ V oscillations (with some high frequency noise). In
  addition, these oscillations are the ``pedestal'' of voltage spikes
  that briefly appear at the end of the top plateaus of the
  square-like oscillations. From this filtering operation, the
amplitude of the spikes could be as high as $\sim 1$ V, suggesting
that during a short period of time many modules are forming a high
field domain at a bias corresponding to the alignment of the upper
lasing state with level 5. With cautiousness one could state that the
deconvoluted traces seem in rather good agreement with some simulated
voltage traces (Fig.~\ref{FigSimOscRegions}(b-d) and 
Figs.~\ref{fig:sim_osc_fft_1}-\ref{fig:sim_osc_fft_3}. 
When the same procedure is performed on oscilloscope
traces recorded for an external bias corresponding to the birth of
oscillations in region III, the spikes of voltage appear in the middle
of the top plateau of the square-like oscillations. 

When the filtering operation is performed on oscilloscope traces
recorded for external biases set near threshold (end of region I,
beginning of region II), the filtered and unfiltered signals have
about the same amplitude oscillations, which suggests that the ringing
effects did not affect very much the measurement. In this case, it
seems that before threshold the oscillations pattern are
triangular-like, and after threshold they become more square-like; an
observation in resonance with simulations displayed in 
Figs.~\ref{fig:sim_osc_fft_0},\ref{fig:sim_osc_fft_1}. Nevertheless, the signal-to-noise and the bandwidth
limitation of the filtering operation prevent a solid confirmation of
these preliminary observations.

To summarize, this deconvolution procedure indicates the severity of
distortion in the measured signal and  emphasizes the importance of
better RF design of voltage detection setup; this would become even
more crucial if chaotic behavior were investigated.

%\bibliography{refs_QuantTrans,add}

\end{document}